\documentclass[review]{elsarticle}

\usepackage{lineno,hyperref}
\modulolinenumbers[5]

\journal{Journal Fluids and Structures}

\usepackage{amsfonts}
\usepackage{amsmath}
\usepackage{graphicx}
\usepackage{epstopdf}
\usepackage{algorithmic}
\usepackage[]{algorithm2e} 
\usepackage{tikz}
\usepackage{subcaption}
\usepackage{floatrow}
\usepackage{wasysym}
\usepackage{multirow}

\usepackage{cleveref}
\crefname{algocfline}{algorithm}{algorithms}
\Crefname{algocfline}{Algorithm}{Algorithms}

\SetKwComment{Comment}{$\triangleright$\ }{}

\usepackage{xargs} 

\usepackage{color, colortbl} 
\definecolor{LightGray}{gray}{0.95}

\usepackage[colorinlistoftodos,prependcaption,textsize=tiny]{todonotes}
\newcommandx{\redc}[2][1=]{\todo[linecolor=red,backgroundcolor=red!25,bordercolor=red,#1]{#2}}
\newcommandx{\bluec}[2][1=]{\todo[linecolor=blue,backgroundcolor=blue!25,bordercolor=blue,#1]{#2}}
\newcommandx{\greenc}[2][1=]{\todo[linecolor=green,backgroundcolor=green!25,bordercolor=green,#1]{#2}}
\newcommandx{\ingray}[2][1=]{\todo[linecolor=gray,backgroundcolor=gray!25,bordercolor=gray,inline,#1]{#2}}
\newcommandx{\ingreen}[2][1=]{\todo[linecolor=green,backgroundcolor=green!25,bordercolor=green, inline,#1]{#2}}
\newcommandx{\inred}[2][1=]{\todo[linecolor=red,backgroundcolor=red!25,bordercolor=red, inline,#1]{#2}}
\newcommandx{\inblue}[2][1=]{\todo[linecolor=blue,backgroundcolor=blue!25,bordercolor=blue, inline,#1]{#2}}

\begin{document}

\begin{frontmatter}

\title{HPC compact quasi-Newton algorithm for interface problems}

\author[BSC]{Alfonso Santiago}
\author[KTH]{Miguel Zavala-Ak\'e}
\author[BSC]{Ricard Borell}
\author[BSC]{Guillaume Houzeaux}
\author[BSC,ELEM]{Mariano V\'azquez}
\ead{mariano.vazquez@bsc.es}

\address[BSC]{Barcelona Supercomputing Center (BSC), Barcelona, Spain.}
\address[ELEM]{ELEM Biotech, Barcelona, Spain}
\address[KTH]{Kungliga Tekniska H\"{o}gskolan (KTH), Stockholm, Sweden.}

\linenumbers

\begin{abstract}
In this work we present a robust interface coupling algorithm called Compact Interface quasi-Newton (CIQN). It is designed for computationally intensive applications using an MPI multi-code partitioned scheme. The algorithm allows to reuse information from previous time steps, feature that has been previously proposed to accelerate convergence.
%
%
%
Through algebraic manipulation, an efficient usage of the computational resources is achieved by: avoiding construction of dense matrices and reduce every multiplication to a matrix-vector product and reusing the computationally expensive loops. This leads to a compact version of the original quasi-Newton algorithm. Altogether with an efficient communication, in this paper we show an efficient scalability up to 4800 cores.
Three examples with qualitatively different dynamics are shown to prove that the algorithm  can efficiently deal with added mass instability and two-field coupled problems. We also show how reusing histories and filtering does not necessarily makes a more robust scheme and, finally, we prove the necessity of this HPC version of the algorithm.
The novelty of this article lies in the HPC focused implementation of the algorithm, detailing how to fuse and combine the composing blocks to obtain an scalable MPI implementation. Such an implementation is mandatory in large scale cases, for which the contact surface cannot be stored in a single computational node, or the number of contact nodes is not negligible compared with the size of the domain. 2020 Elsevier.
This manuscript version is made available under the CC-BY-NC-ND 4.0 license http://creativecommons.org/licenses/by-nc-nd/4.0/
\end{abstract}

\begin{keyword}
coupling scheme \sep fluid-structure interaction \sep high performance computing \sep partitioned scheme 
\MSC[2010] 68U20 \sep 00A72 \sep 68Q85 \sep 65-04 \sep 74F10 
\end{keyword}

\end{frontmatter}



\section{Introduction}
\label{sec:introduction}
Interface problems like, solid-solid contact, fluid-structure interaction (FSI) or heat transfer gained great attention in the last decades due to the broad range of applications in aerospace industry, manufacturing, wind energy production or biomechanics. These problems can be mathematically treated using heterogeneous domain decomposition methods \cite{Deparis2006}. In each subdomain the problems are defined with their own Neumann and Dirichlet boundary conditions, including the boundary condition at the contact surface. From the algorithmic point of view, the problem can be attacked with the monolithic or the partitioned scheme. On the former, using and \textit{ad-hoc} solver, one matrix is build including the degrees of freedom for both the fluid and solid \cite{Gee2010,Crosetto2011,Hron2006a}. On the latter, fluid and the solid are computed independently as black-box solvers, exchanging the quantities of interest at certain synchronisation points of the workflow \cite{Degroote2009,Matthies2003,Habchi2013,Radtke2016}. Both strategies have advantages and drawbacks. Monolithic schemes have less numerical instabilities, but leads to a linear system hard to preconditionate and require to design an specific solver from scratch for every pair of coupled problems \cite{Gee2010,Hron2006a,Verdugo2015a,Badia2008a}. The partitioned scheme allows code reusing, but require convergence iterations at each time step \cite{Degroote2009,Matthies2003,Habchi2013,Radtke2016}.

The algorithm presented in this work is based on the interface quasi-Newton method (IQN) \cite{Degroote2009}, improved and extended with a special care in the parallel implementation. The proposed scheme is implemented in Alya \cite{Houzeaux2009, Houzeaux2011, Casoni2014}, the BSC's in-house tool for multiphysics problems. The uncoupled physics solvers in Alya have an almost linear scalability proven up to a hundred thousand cores \cite{Vazquez2014}. Following a black-box multi-code strategy, the coupled problems are solved by executing two different MPI-based parallel instances of Alya which interchange data in the contact surface.
The goal of this work is to develop both an accurate and efficient version of the interface quasi Newton algorithm. There are three main motivations to do so. Firstly, we look for a coupling strategy that can deal with large-scale problems on both sides of the coupled problem. Secondly, we need a coupling algorithm that is able to robustly tackle the main issue on the fluid-structure interaction (FSI) problem, namely added mass instability \cite{Forster2007,Causin2005}. And, finally, the algorithm should be able to deal with more than one coupled interface at the same time (n-field coupling) \cite{Bungartz2015}.
When modelling biomechanics, the last two conditions are mandatory, as the densities of the tissues and fluids are similar and multiple cavities are interconnected. A parallel formulation of the coupling algorithm has  not been shown in the past, but it is a requirement for massively parallel applications so it does not becomes the bottleneck. Previous publications \cite{Degroote2009,Matthies2003,Habchi2013,Radtke2016,uekermann2016partitioned,Haelterman2016} assume a negligible cost of the coupling algorithm. This is based on the fact that in some cases the coupled surface might be small compared with the solved volumes. This assumption eases the software development as the code can be written as serial. This hypothesis might be true in some cases, but not in biomechanical applications where the geometrical complexity of the biological structures enormously increases the area of the contact surface. Although it is true that the convergence acceleration require less operations than the solvers in the Piccard iteration, the interface quasi-Newton algorithm requires a large number of matrix-matrix multiplications. This is intractable for large problems that must run in distributed memory systems where the interface cannot be stored in a single shared-memory node, so an efficient parallelisation is mandatory.

This work is organised as follows. \Cref{sec:methods} contains the mathematical development for the algorithm, detailing the included improvements and the parallelisation strategy. \Cref{sec:experiments} shows two experiments with remarkably different dynamics and a scalability test. Conclusions can be found in \cref{sec:conclusion_and_fw}.

%



\section{Material and Methods}
\label{sec:methods}

In this section, a compact version of the Interface quasi-Newton algorithm is presented. \Cref{subsec:methods::algorithm} shows the original algorithm, and \cref{subsec:methods::improvements} the included improvements. The QR decomposition, a critical step in the algorithm, is thoroughly detailed in \cref{subsec:methods::QR} and the parallelisation strategy in \cref{subsec:CompactIQN}. An explanation of the used Einstein index notation convention can be found in \ref{sec:appx:index_convention}.

\subsection{Problem setting}
\label{subsec:methods::problemsetting}

In this work, we focus in surface problems that can be stated as $d_\alpha^{I+1}=S(f_\alpha)$ and $f_\alpha^{I+1}=F(d_{\alpha})$. Each form represents the numerical result of a physical problem and  $d_{\alpha}$ and $f_\alpha$ are the unknowns at the interface $\Gamma_c$. This can also be written as the fixed point equation $d_\alpha^{I+1}=\mathrm{S}(\mathrm{F}(d_\alpha))$, or in a generic manner:
\begin{equation}
\widetilde{x}_\alpha^{I+1}=\mathrm{H}(x_\alpha),
\label{eqn:fixed_point}
\end{equation}
where $\mathrm{H}(x_\alpha)$ condenses both solvers. The parallel solvers $S(f_\alpha)$ and $F(d_{\alpha})$ can be executed either one after the other in a block-sequential manner (Gauss-Seidel) or at the same time in a block-parallel manner (Jacobi) \cite{Mehl2016}. While the former is less computationally efficient, it improves convergence of the iterative solver, and therefore will be the used scheme. Performance can be improved with a convergence acceleration algorithm. Examples of them can be found in \cite{Degroote2013}. In the following section we will develop a high-performance version of an interface quasi-Newton algorithm.


%

\subsection{General Overview of the Algorithm}
\label{subsec:methods::algorithm}

The first implementation of the Interface Quasi Newton (IQN) algorithm is described in \cite{Degroote2009}. Distinctly to other quasi-Newton schemes, in the IQN the Jacobian is approximated by a field defined in the contact surface and depending on the local residual variation over a given number of iterations \cite{Scheufele2015}. The residual of \cref{eqn:fixed_point} can be defined as $r_\alpha=H(x_\alpha) - x_\alpha = \widetilde{x}_\alpha - x_\alpha$. For each time step, the problem is converged when $r_\alpha=0$. If the Jacobian ${\partial r_\alpha}/{\partial x_\beta}$ is known, the increment of the variable ${x}_\alpha$ can be computed as:
\begin{equation}
\frac{\partial r_\alpha}{\partial x_\beta}\Delta x_\beta=-r_\alpha.
\label{eqn:derivative_residue}
\end{equation}
Therefore, computing the next iterate as $x_\alpha^{I+1}=x_\alpha+\Delta x_\alpha$. Generally the exact Jacobian cannot be computed or it is computationally expensive to do so. This quasi-Newton scheme provides a method to obtain an approximation of the inverse Jacobian. The multi-secant equation for the inverse Jacobian reads:
\begin{equation}
\left(\frac{\partial r_\alpha}{\partial x_\beta}\right)^{-1}V_{\alpha i}\approx W_{\beta i},
\end{equation}
where:
%
%

\begin{align}
V_{\alpha i}&=\left[\Delta r^{1}_\alpha ,\Delta r^{2}_\alpha, ... , \Delta r^q_\alpha \right]
&\text{\hspace{0.25cm}with\hspace{0.25cm}}&
&\Delta r_\alpha^{I}&= r_\alpha^{I+1} - r_\alpha^I  \label{eqn:residincr} \\
W_{\alpha i}&=\left[\Delta \widetilde{x}_\alpha^{1} ,\Delta \widetilde{x}_\alpha^{2}, ... , \Delta \widetilde{x}_\alpha^q \right]
&\text{\hspace{0.25cm}with\hspace{0.25cm}}&
&\Delta \widetilde{x}^{I}_\alpha&= \widetilde{x}_\alpha^{I+1} - \widetilde{x}_\alpha^I . \label{eqn:valincr}
\end{align}

where $r_\alpha^{I+1}$ and $\widetilde{x}_\alpha^{I+1}$ and $r_\alpha^{I}$ and $\widetilde{x}_\alpha^{I}$ are the current and past values respectively. $V_{\alpha i},W_{\alpha i}\in \mathbb{R} ^{p\times q}$, where $p$ is the number of contact degrees of freedom and $q$ is the number of saved non-zero iterations where, generally, $p>>q$. Note that the newest values are stored at the left side of the built matrix, while the older values are moved to the right. The residual increment of the current iteration is approximated as a linear combination of the previous residuals increments:
\begin{equation}
\label{eqn:deltar}
\Delta r_\alpha = V_{\alpha i} \lambda_{i},
\end{equation}
where $\lambda_i \in \mathbb{R} ^{q\times1}$ is the solution of the optimisation problem $\parallel\Delta r_\alpha - V_{\alpha i}\lambda_i\parallel$ described in \cite{Vierendeels2007}. To obtain $\lambda_i$, the matrix $V_{\alpha i}$ is decomposed in an orthogonal matrix $Q_{\alpha \beta} \in \mathbb{R} ^{p\times p}$ and an upper triangular $U_{\alpha i}  \in \mathbb{R}^{p\times p}$ with a QR decomposition:
\begin{equation}
V_{\alpha i}=Q_{\alpha \beta}U_{\beta i}.
\label{eqn:VQU}
\end{equation}

As $U_{\alpha i}$ is upper triangular, only its first $q$ rows are different from zero. With this information we can build  a modified QR decomposition with $U_{i j} \in \mathbb{R} ^{q\times q}$ and  $Q_{\alpha i} \in \mathbb{R} ^{p\times q}$ such that:
\begin{equation}
{V_{\alpha i}}={Q}_{\alpha k}{U}_{k i},
\label{eqn:ecoVQU}
\end{equation}
reducing the amount of memory and computing effort required, as described ahead in \cref{subsec:CompactIQN}. After this decomposition, the vector $\lambda_i$ can be obtained by backsubstitution of the upper triangular matrix $U_{i j}$:
\begin{equation}
U_{i j}\lambda_j={Q_{\alpha i}} \Delta r_\alpha .
\end{equation}
As $Q_{\alpha i}$ is orthogonal, the inverse is equal to the transpose, avoiding the inversion of this matrix. Also, as $\Delta r_\alpha=r_\alpha^{I-1}-r_\alpha$ and the objective is to get $\Delta r_\alpha=0_\alpha-r_\alpha$, we can say:
\begin{equation}
U_{i j}\lambda_j={-Q_{\alpha i}} r_\alpha .
\label{eqn:backsualpha}
\end{equation}
Once $\lambda_i$ is computed, the increment of the unknown $\Delta x_\alpha$ can be computed as $\Delta x_\alpha=W_{\alpha i} \lambda_i$, and the update of the unknown as:
\begin{equation}
x^{I+1}_\alpha=\widetilde{x}_\alpha+W_{\alpha i}\lambda_i.
\label{eqn:update}
\end{equation}
%

The scheme is summarised in \cref{algorithm:iqn}. For each time iteration, an initial guess and residue are computed. As the proposed algorithm requires increments, a first step with fixed relaxation $\omega_0$ is required. After, IQN loop continues until convergence is achieved.
\vspace{0.5cm}
\begin{algorithm}[H]
	\DontPrintSemicolon
	\nl For each time step, solve:\;
	\nl $x_\alpha^0=x_\alpha^{ini}$\;
	\nl $\widetilde{x}_\alpha^0=H(x_\alpha^0)$\;
	\nl $r_\alpha^0=\widetilde{x}_\alpha^0 -x_\alpha^0$\;
	\nl update $x_\alpha^{1}= \widetilde{x}_\alpha^0 + \omega_0 r_\alpha^0$\;
	\While{problem not converged}{
	\nl 	$\widetilde{x}_\alpha=H(x_\alpha)$\;
	\nl 	$r_\alpha=\widetilde{x}_\alpha -x_\alpha$\;
		
	\nl 	build $V_{\alpha i}=\left[\Delta r_\alpha^{I-1}, ... , \Delta r_\alpha^{0} \right]$; with $\Delta r_\alpha^{I}= r_\alpha^I - r_\alpha$\;
	\nl 	build $W_{\alpha i}=\left[\Delta \widetilde{x}_\alpha^{I-1}, ... , \Delta \widetilde{x}_\alpha^{0} \right]$; with $\Delta \widetilde{x}_\alpha^{I}= \widetilde{x}_\alpha^I - \widetilde{x}_\alpha$\;
		
	\nl 	decompose $V_{\alpha j}=Q_{\alpha i}U_{i j}$ (by QR decomposition) \;\label{algorithm:iqn:linerq} 
		
	\nl 	solve $U_{i j} \lambda_j =-{Q_{\alpha i}} r_\alpha$ \;
		
	\nl 	update $x_\alpha^{I+1}= \widetilde{x}_\alpha + W_{\alpha i} \lambda_i$
		
	}
	\vspace{0.2cm}
	\caption{Interface quasi-Newton algorithm overview.\label{algorithm:iqn}}
\end{algorithm}
\vspace{0.5cm}
%

\subsection{Improvements on the original scheme}
\label{subsec:methods::improvements}
It has been proposed \cite{Haelterman2016} that adding information from iterations from the previous time steps into matrices $V_{\alpha i}$ and $W_{\alpha i}$ improve the convergence properties of the algorithm. To do so, we redefine  matrices \ref{eqn:residincr},\ref{eqn:valincr}, with information of the iterations from previous time steps:
\begin{align}
V_{\alpha i}&=\left[V_{\alpha i}^1,V_{\alpha i}^2, ... , V_{\alpha i}^T, \right]
&\text{\hspace{0.25cm}with\hspace{0.25cm}}&
& V_{\alpha i}^t  \quad\text{as \cref{eqn:residincr} } \label{eqn:Tresidincr} \\
W_{\alpha i}&=\left[W_{\alpha i}^1,W_{\alpha i}^2, ... , W_{\alpha i}^T, \right]
&\text{\hspace{0.25cm}with\hspace{0.25cm}}&
& W_{\alpha i}^t  \quad\text{as \cref{eqn:valincr}, } \label{eqn:Tvalincr}
\end{align}
where $t$ ranges from the current processed time step to the last saved time step, $T$. Note that $V_{\alpha i},W_{\alpha i}\in \mathbb{R} ^{p\times q}$, but now $q$ is the number of saved non-zero iterations from the current and past time steps. Including this information increases the probability that the columns in $V_{\alpha i}$ are linearly dependent, rendering the QR decomposition unstable. Different filtering techniques have been proposed \cite{Haelterman2016} to remove these columns, but all of them require building dense intermediate matrices, or even finding every associated eigenvalue \cite{Bogaers2014}, with its associated expensive computational cost. Moreover there is not a clearly better filtering technique \cite{uekermann2016partitioned,Haelterman2016}. This is why we choose to reuse the simple and paralellised incomplete QR decomposition developed in this work to check the linear dependency of the columns of $V_{\alpha i}$. If $\left|U_{ii}\right|<\epsilon\left|\left|U\right|\right|_2$, where $U_{ij}$ is the upper triangular matrix and $\epsilon$ a parameter, the $i$-th column is deleted from $V_{\alpha i}$ and $W_{\alpha i}$. The column deleted might correspond to the current processed time step (sub-matrix $V_{\alpha i}^1$) or any other column corresponding to any other time step (sub-matrix $V_{\alpha i}^j$). This requires re-stacking the non-zero columns to obtain again a dense set of matrices.

\subsection{QR decomposition}
\label{subsec:methods::QR}

A critical step is the QR decomposition (\cref{algorithm:iqn:linerq} in \cref{algorithm:iqn}) due to the numerous matrix-matrix products involved in it. In this section the QR decomposition will be explained and through algebraic manipulation these matrix products will be simplified in the following section. The goal of the QR decomposition is to obtain  the orthogonal and the upper triangular matrices $Q_{\alpha \beta}$ and $U_{\alpha i}$, with the following shape:
\begin{align}
	Q_{\alpha \epsilon} &=\ ^1B_{\alpha \beta}\ ^2B_{\beta \gamma} ... \ ^qB_{\gamma \epsilon} \label{eqn:obtainq} \\
	U_{\alpha i} &=\ ^qB_{\alpha \beta}...\ ^2B_{\beta \gamma}\ ^1B_{\gamma \epsilon}V_{\epsilon i}, \label{eqn:UisQV} 
\end{align}
where $\ B_{\alpha \beta} \in \mathbb{R}^{p\times p}$ are dense intermediate matrices obtained during the iterative decomposition. At each iteration, the matrix $V_{\alpha i}$ is processed column by column. We use a left superscript to identify 
the corresponding iteration of the QR algorithm but, for easiness on the reading, we avoid using any other time or coupling iteration superscripts. $V_{\alpha i}$ can be considered as a set of  $q$ ordered vectors:
\begin{equation}
\ ^1V_{\alpha i} =%
\begin{bmatrix} %
\begin{bmatrix}
v_{1 1} \\ %
v_{2 1} \\ %
\vdots     %
\end{bmatrix} & %
\begin{bmatrix}
v_{1 2} \\ %
v_{2 2} \\ %
\vdots     %
\end{bmatrix} & %
  \cdots  &%
  \begin{bmatrix}
  v_{1 q} \\ %
  v_{2 q} \\ %
  \vdots     %
  \end{bmatrix} & %
\end{bmatrix}
= \left[v_{\alpha 1},v_{\alpha 2},\cdots,v_{\alpha q}\right]. 
\end{equation}
The algorithm, iteratively makes each column orthogonal to each other column in the matrix. It starts iteration $j$ with a matrix $\ ^jV_{ \alpha i}$ obtained with data from iteration $j$-1. To decompose the $j$-th column of $^jV_{\alpha i}$, a unitary vector $u_\alpha$ has to be built:
\begin{equation}
{u}_\alpha =\frac{{n}_\alpha}{ \lVert{n}_\alpha \lVert} \quad \textrm{with,} \quad {n}_\alpha ={v}_\alpha- \lVert{v}_\alpha\lVert\ ^j{e}_\alpha, \label{eqn:normailze_u}
\end{equation}
where $v_\alpha$ is the column to decompose and $\ ^je_\alpha$ is a unitary vector with $j$-th position equal to 1 and to 0 otherwise. Then,
\begin{equation}
^jB^*_{\alpha \beta}=\delta_{\alpha \beta}-2{u}_\alpha{{u}_\beta} 
\label{eqn:matrixQ}
\end{equation}
is the so called Householder matrix, and $\delta_{\alpha \beta}$ is the identity matrix. If $^jV_{\alpha i}$  is premultiplied by $^jB^*_{\alpha \beta}$, a new matrix $^jB^*_{\alpha \beta}\ ^jV_{\beta i}$ is obtained: 
\begin{equation}
\ ^1B^*_{\alpha \beta}\ ^1V_{\alpha i} = %
\begin{bmatrix} %
 \lVert {v}_{\alpha 1}\lVert		  & \cdots &  \cdots  &  \cdots\\ %
 0													&  		 	 &             &       	\\ %
 \vdots											 &  		  &  $$^2V_{\beta j}$$	  &   		\\ %
 0 													&  			 &  		    & 				& %
 \end{bmatrix}\label{eqn::mtxbv}
\end{equation}
Matrix \ref{eqn::mtxbv} is upper triangular in the first $j$ columns; and dense everywhere else. A new submatrix $^{j+1}V_{\beta j}$ is therefore defined after erasing the first column and row. This process can be repeated until the initial matrix becomes upper triangular.

Once the algorithm is computed for every column on $^1V_{\alpha i}$, a set of $q$ gradually smaller matrices ${^1B^*_{\alpha i}}\in \mathbb{R}^{p\times p}$, ${^2B^*_{\alpha i}}\in \mathbb{R}^{p-1\times p-1}$ ... ${^jB^*_{\alpha i}}\in \mathbb{R}^{p-(j-1)\times p-(j-1)}$  ... ${^qB^*_{\alpha i}}\in \mathbb{R}^{1\times 1}$ are obtained. To properly compute the $j$-ith iteration of \cref{eqn::mtxbv}, matrices $^jB_{\alpha i}$ are completed with the identity:
\begin{equation}
 {^jB}_{\alpha i} =
\begin{bmatrix}
$$ I_{i j}$$  &   0 \\ %
0								 &  $$ {^jB}^*_{\alpha i} $$	%
\end{bmatrix}
\end{equation}
where $ I_{i j}\in \mathbb{R}^{j-1\times j-1}$. Finally, through \cref{eqn:obtainq,eqn:UisQV} the matrices  $U_{i j}$ and $Q_{\alpha i}$ are obtained. The process is described in \cref{algorithm:qrdecomposition}.

\begin{algorithm}[H]
	\vspace{0.2cm}
	\DontPrintSemicolon
	\nl $^1V_{\alpha i}=V_{\alpha i}$\;
	\For{j=1...q}{
	\nl 	choose ${v}_\alpha=\ ^jV_{\alpha i}$ with $\alpha=j...p$ and $i=j$\;
	\nl 	${n}_\alpha=\ {v}_\alpha- \lVert{v}_\alpha\lVert{e}_\alpha$\;
	\nl 	${u}_\alpha={\ {n}_\alpha}/{ \lVert {n}_\alpha \lVert}$\;
	\nl 	$B^*_{\alpha \beta}=I-2\ {u}_\alpha{\ {u}_\beta} $\;
	\nl 	$^{j+1}V_{\alpha i}=\ B_{\alpha \beta}\ V_{\beta i}$
	}
	\nl $Q_{\alpha \epsilon}=\ ^1B_{\alpha \beta}\ ^2B_{\beta \gamma} ...\ ^qB_{\gamma \epsilon}$\;
	\nl $U_{\alpha i}=\ ^qB_{\alpha \beta} \ ^{q-1}B_{\beta \gamma} ...\ ^1B_{\gamma \epsilon}V_{\epsilon i} $\;
	\vspace{0.2cm}
	\caption{overview of the QR decomposition algorithm.\label{algorithm:qrdecomposition}}
\end{algorithm}
\vspace{0.5cm}

\subsection{Paralell compact IQN}
\label{subsec:CompactIQN}

\begin{figure}[h]
	\centering
	\includegraphics[width=0.4\textwidth]{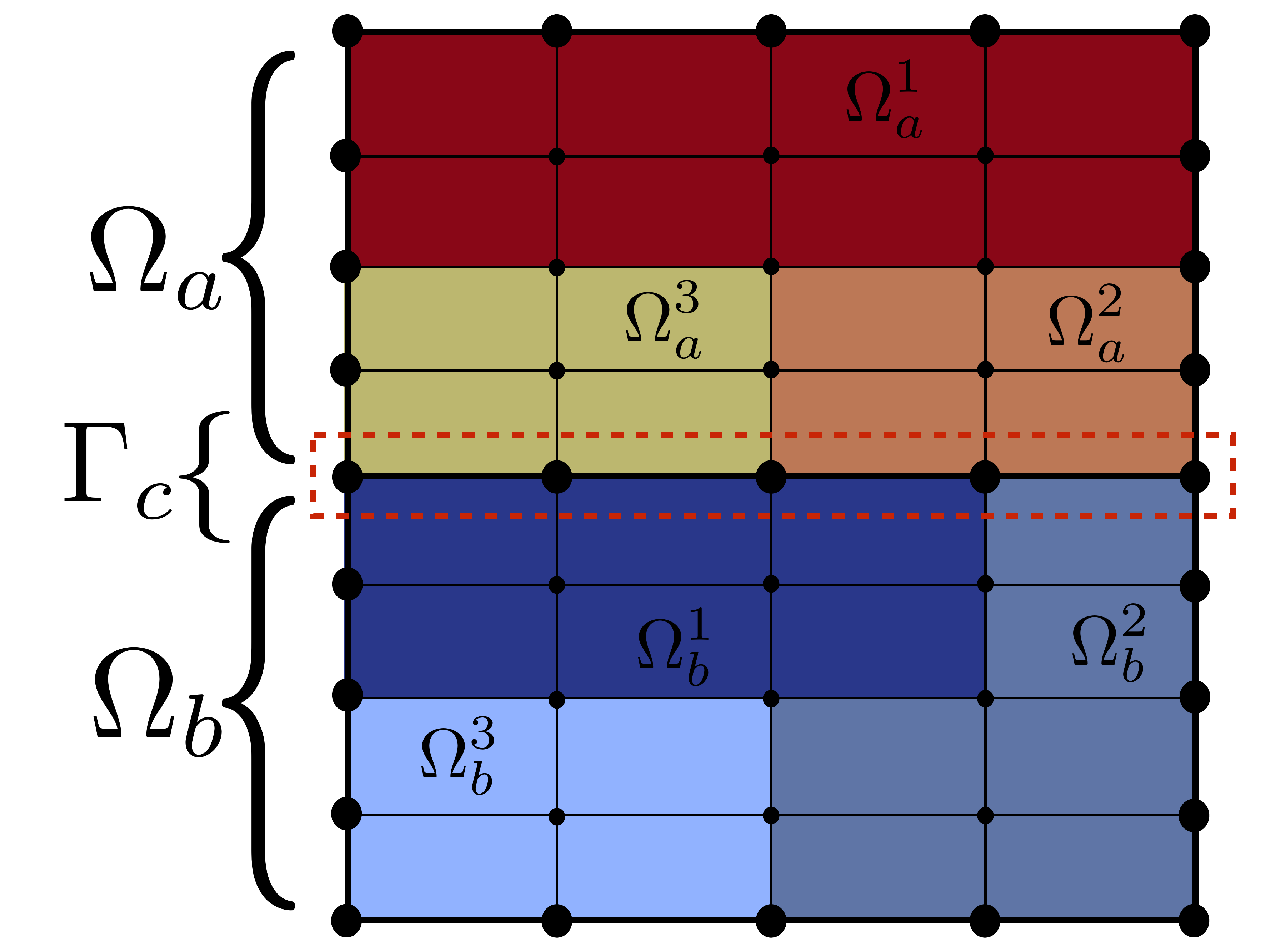}
	\caption{Physical subdomains $\Omega_a$ and $\Omega_b$ in contact by the wet surface $\Gamma_c$. Each physical subdomain is subdivided in three computational subdomains (partitions). The wet surface, and therefore the vectors in matrix $V_{\alpha i}$ can be distributed along several partitions. \label{fig:parallpartition}}
\end{figure}

The distributed memory parallelisation of Alya is based on a domain decomposition \cite{HOUZEAUX2018216}, a mesh partition is carried out \cite{BORRELL2018264}, and each partition is assigned to a MPI-process. The mesh partitioner divides the mesh minimising the area between subdomains but without any requirements on the contact surface $\Gamma_c$ (see \cref{fig:parallpartition}). Therefore, the nodes in $\Gamma_c$ will be distributed among the MPI tasks and so the increment matrix $V_{\alpha i}$.
In order to process $V_{\alpha i}$ in parallel, we look for the subdomain with the largest number of contact nodes, the now called ``leader'' partition. After, $V_{\alpha i}$ and $W_{\alpha i}$ are renumbered so the first rows correspond to the leader partition, so the backsubstitution is only executed there. 

To improve computing and memory cost, we propose some modifications for the base algorithm in \cref{subsec:methods::QR}. As $^jB_{\alpha \beta}$ is obtained by \cref{eqn:matrixQ}, the product $^jB_{\alpha \beta}\ ^jV_{\beta i}$ can be expanded as:
\begin{align}
	^jB_{\alpha \beta}\ ^jV_{\beta i}=(\delta_{\alpha \beta}-2{u_\alpha}{u_\beta})\ ^jV_{\beta i} = \ ^jV_{\alpha i}-2{u_\alpha}{u_\beta}\ ^jV_{\beta i}. \label{eqn:expansionA}
\end{align}
So, instead of computing and storing $^jB_{\alpha \beta} \in \mathbb{R}^{p\times p}$ for each iteration $j$, we store the vectors $u_\alpha\in \mathbb{R}^{p}$ for the $q$ iterations. Expanding \cref{eqn:expansionA}:
\begin{align}
(\delta_{\alpha \beta}-2{u_\alpha}{u_\beta})\ ^jV_{\beta i} =\delta_{\alpha \beta}\ ^jV_{\beta i}-2{u_\alpha}{u_\beta}\ ^jV_{\beta i} = \ ^jV_{\alpha i}-2{u_\alpha}\left({u_\beta}\ ^jV_{\beta i}\right). \label{eqn:expansion}
\end{align}
So the parallel matrix-vector product ${u_\alpha}\ ^jV_{\alpha i}$ can be computed first, then compute ${u_\alpha}({u_\beta}\ ^jV_{\beta i})$ and finally subtract $I_{\alpha \beta}\ ^jV_{\beta i}-2{u_\alpha u_\beta}\ ^jV_{\beta i}$. To compute \cref{eqn:UisQV} we proceed similarly, but starting with $(\delta_{\alpha \beta}-2{\ ^ 1 u_\alpha}{\ ^ 1 u_\beta})V_{\beta i}$ followed by the premultiplication of the matrices $(\delta_{\alpha \beta}-2{u_\alpha}{u_\beta})$ with the same technique as described here.

Similarly, we can avoid the construction of the dense matrix $Q_{\alpha \beta}\in \mathbb{R}^{p\times p}$, used in the backsubstitution (see \cref{eqn:backsualpha}). Vector $-Q_{\alpha i} {r}_\alpha$ can be computed  with a strategy similar to \cref{eqn:expansionA}. The difference is that ${Q_{\alpha \epsilon}}=\ ^1B_{\alpha \beta} \cdots \ ^qB_{\gamma \epsilon}$, so after computing ${\ ^q B_{\alpha i}} {r}_\alpha$ as:
 \begin{align}
 	{\ ^q B_{\alpha \beta}} {r}_\alpha=(\delta_{\alpha \beta}-2\ ^q{u}_\alpha{\ ^q{u}_\beta}){r}_\beta = r_\beta-2\ ^q{u}_\alpha{\ ^q{u}_\beta} {r}_\beta, \label{eqn:expansion2A}
 \end{align}
the rest of the matrices $ B_{\alpha \beta} =\delta_{\alpha \beta}-2{u_\alpha}{u_\beta}$ are premultiplied. The first multiplication, (\cref{eqn:expansion2A}) can be expanded as:
\begin{align}
(\delta_{\alpha \beta}-2\ ^q{u}_\alpha{\ ^q{u}_\beta}){r}_\beta = {\delta_{\alpha \beta}}r_\beta-2\ ^q{u}_\alpha{\ ^q{u}_\beta} {r}_\beta = r_\alpha-2\ ^q{u}_\alpha\left({\ ^q{u}_\beta} {r}_\beta\right) \label{eqn:expansion2B}
\end{align}
The product ${\ ^q{u}_\beta} {r}_\beta$ is firstly computed and then $ r_\alpha-2\ ^q{u}_\alpha({\ ^q{u}_\beta  r_\beta})$. The resulting vector is multiplied by $ \left(\delta_{\alpha \beta} - 2{\ ^{(q-1)} u_\alpha}{\ ^{(q-1)} u_\beta}\right)$ and followed by every matrix $\left(\delta_{\alpha \beta} - 2{\ ^{j} u_\alpha}{\ ^{j} u_\beta}\right)$ up to  $\left(\delta_{\alpha \beta}-2{\ ^{1} u_\alpha}{\ ^{1} u_\beta}\right)$. In this way, matrices $^q B_{\alpha i}$ are never completely computed.

The resulting algorithm has as input the $V_{\alpha i}$ matrix and the residuals vectors to operate in the backsubstitution (see \cref{eqn:backsualpha}), and the output will be the coefficient vector ${\alpha}_i$. As the boundaries between the IQN and the QR algorithms can't be identified anymore, we refer to the developed algorithm as Compact IQN (CIQN). Our main motivation is that a complete QR decomposition would be prohibitive in large cases as a dense $Q$ orthogonal matrix would be extremely expensive to compute and store. The proposed algorithm is a collection of matrix-vector and vector-vector products restricted to the contact. An efficient parallelisation requires a proper point-to-point MPI communication on the modified IQN and QR algorithms. The whole sequence of steps is described in \cref{algorithm:ciqn23}.

%
\begin{algorithm}[H]
	\SetAlgoLined  
	\DontPrintSemicolon
	\nl Chose leader partition\;
	 \While(\Comment*[f]{time loop}){not the last time step}{
	\nl $x_\alpha^0=x_\alpha^{ini}$\;
	\nl $\widetilde{x}_\alpha^0=H(x_\alpha^0)$\;
	\nl $r_\alpha^0=\widetilde{x}_\alpha^0 -x_\alpha^0$\;
	\nl update $x_\alpha^{1}= \widetilde{x}_\alpha^0 + \omega_0 r_\alpha^0$\;
	 \While{problem not converged}
		{
	\nl 	$\widetilde{x}_\alpha=H(x_\alpha)$\;
	\nl 	$r_\alpha=\widetilde{x}_\alpha -x_\alpha$\;
	\nl 	build $V_{\alpha i}$ and $W_{\alpha i}$ as \cref{eqn:Tresidincr,eqn:Tvalincr} \;
	
	\nl 	$\ ^1V_{\alpha i}=V_{\alpha i}$\;
		
		\For(\Comment*[f]{QR decomposition loop}){j=1...q}
			{ 
			\lIf{$j>1$}{
	 			$^{j+1}V_{\alpha i}= \ ^1V_{\beta i}-2{u_\alpha}{u_\beta} \ ^1V_{\beta i}$ as \cref{eqn:expansion}}
			\eIf{I am the leader}{
		\nl 		${v}_\alpha=\ ^jV_{\alpha i}$ with $\alpha=j...p$ and $i=j$\;
				}{
		\nl 		${v}_\alpha=\ ^jV_{\alpha i}$ with $\alpha=1...p$ and $i=j$\;
				}
		\nl 		parallel compute $\lVert{v}_\alpha\lVert$\;
	
				\lIf{I am leader}{
		 		${n}_\alpha=\ {v}_\alpha- \lVert{v}_\alpha\lVert{\ ^je}_\alpha$}
	
		\nl 		parallel compute ${ \lVert {n}_\alpha \lVert}$\;
	
		\nl 		${u}_\alpha={ {n}_\alpha}/{ \lVert {n}_\alpha \lVert}$\;
			} 

	\nl $U_{i j} =\left(\delta_{i \gamma}-2{\ ^q u_i}{\ ^q u_\gamma}\right)\cdots \left( \ ^1V_{\alpha j}-2{\ ^1 u_\alpha}{\ ^1 u_\beta}\ ^1V_{\beta j}\right)$ (as \cref{eqn:expansion})\;
    \lIf{$|U_{jj}|<\epsilon ||U||_2$}{remove column $j$ and restart {\fontfamily{qcr}\selectfont QR} loop}
	\nl $-Q_{\alpha i} {r}_\alpha = -\left(\delta_{i \gamma}-2{\ ^1 u_i}{\ ^1 u_\gamma}\right)\cdots\left( r_\beta -2{\ ^q u_\beta}{\ ^q u_\alpha}r_\alpha \right) $ (as \cref{eqn:expansion2B})\;


			\lIf{I am leader} {backsubstitute $U_{i j}{\lambda_j}=-{Q_{\alpha i}} {r_\alpha}$ }
			
			\nl $x_\alpha^{I+1}= \widetilde{x}_\alpha + W_{\alpha i} \lambda_i$ \Comment*[r]{variable update}
		
		} 
	}
	
	\caption{Compact Interface quasi-Newton algorithm. \label{algorithm:ciqn23}}
\end{algorithm}
%


\subsection{Physics of solved cases}
\label{subsec:methods::physics}

The algorithm in this work has been developed generically for any interface problem. Although that, the main interest of the authors is Fluid-Structure Interaction (FSI), we briefly describe the governing equations. The Newtonian fluid is modelled with incompressible Navier-Stokes equations using an Arbitrary Lagrangian-Eulerian (ALE) formulation:
\begin{align}
	\rho^f\frac{\partial{u_i}}{\partial t} +\rho^f\left(u_j-u_j^m\right)\frac{\partial{u_i}}{\partial x_j} + \frac{\partial}{\partial x_j}\left[+p\delta _{ij} -\mu\left(\frac{\partial u_i}{\partial x_j} +\frac{\partial u_j}{\partial x_i}\right)\right] &=+\rho^f f_i \label{eqn:momentum}\\
	\frac{\partial u_i}{\partial x_i} &= 0,
\end{align}
where $\mu$ is the viscosity of the fluid, $\rho^f$ the density, ${u}_i$ the velocity, $p$ is the mechanical pressure, $f_i$ the force term and ${u}^m_j$ is the fluid domain velocity. The numerical model is based on the Finite Element Method, using the Variational Multiscale\cite{Houzeaux2008}.
For the Arbitrary Lagrangian-Eulerian (ALE) formulation, the technique used is proposed in \cite{Calderer2010}. Mesh movement is solved through a Laplacian equation
\begin{equation}
  \frac{\partial}{\partial x_j} \left(\left[1+\alpha^e\right] \frac{\partial b_i }{\partial x_j} \right) = 0,
\end{equation}
where $b_i$ are the components of the displacement in each point for the domain. The factor $\alpha$ is a diffusive term that, once discretised, controls the mesh distortion. ALE boundary conditions at the contact surface is set through the nodal displacement from solid mechanics problem. 

Solid mechanics is solved following a transient scheme and using a total Lagrangian formulation in finite strains \cite{Casoni2014}. The displacement $d_i$ form of the linear momentum balance can be modelled as:
\begin{equation}
\rho^s \frac{\partial^2 {d}_i}{\partial^2 t} = \frac{\partial P_{iJ}}{\partial X_J}+\rho^s {B}_i,
\end{equation}
where $\rho^s$ is the initial density of the body, ${B}_i$ represents the body forces and $P_{iJ}$ is the nominal stress tensor. Solid mechanics boundary conditions at the contact surface is set through the nodal forces from the fluid mechanics problem.

Let us label ``$\mathrm{CFD}$''and ``$\mathrm{CSM}$'' the fluid and solid sides of a coupled FSI problem. At the contact surface, displacements and normal stresses must be continuous:
\begin{align}
\ ^\mathrm{CFD}{d}^{\Gamma_c}_i &= \ ^\mathrm{CSM} {d}^{\Gamma_c}_i \label{eqn:displcont}\\
n_i\ ^\mathrm{CFD}{\sigma}^{\Gamma_c}_{ij} &= n_i\ ^\mathrm{CSM} {\sigma}^{\Gamma_c}_{ij}\label{eqn:stresscont},
\end{align}
where $\ ^\mathrm{CFD}{d}^{\Gamma_c}_i$ and $\ ^\mathrm{CSM} {d}^{\Gamma_c}_i$ are the deformation in the contact boundary for the fluid and for the solid respectively; and $n_i\ ^\mathrm{CSM} {\sigma}^{\Gamma_c}_{ij}$ and $n_i\ ^\mathrm{CFD}{\sigma}^{\Gamma_c}_{ij}$ are the normal stresses in the contact boundary.

A typical behaviour of the developed algorithm is shown in \ref{subsec:appx:typic} for an FSI case. Solver iteration are decomposed in the different parts: Momentum, continuity and ALE for the fluid domain, and displacement for the solid domain.



\section{Results and discussion}
\label{sec:experiments}

In this section we present tree cases.  Problems in \cref{subsec:exp:tube,subsec:exp:angrybird} were chosen to show the difference of the behaviour of the algorithm with different dynamics on the physics, while problem in \cref{subsec:exp:scalability} is an scalability test. Every problem in this section can be executed relaxing the force or the displacement. The best (less average iterations) scheme for each case is shown in this section, the rest in \ref{sec:appx:results}. For each case, a sensitivity analysis is executed for the number of past saved time steps (histories), iterations on each time step (ranking) and $\epsilon$. Also, as a reference, the number of iterations is compared against the popular Aitken  algorithm. For this comparison algorithm, results are shown as \textit{(e.g.)} 17.76 (sd=2.91) where the first figure indicates the mean and the second figure the standard deviation (sd). All cases are executed in Marenostrum IV supercomputer.

\subsection{Algorithm validation}
\label{subsec:validation}

\begin{figure}[h]
	\centering
	\includegraphics[width=0.85\textwidth]{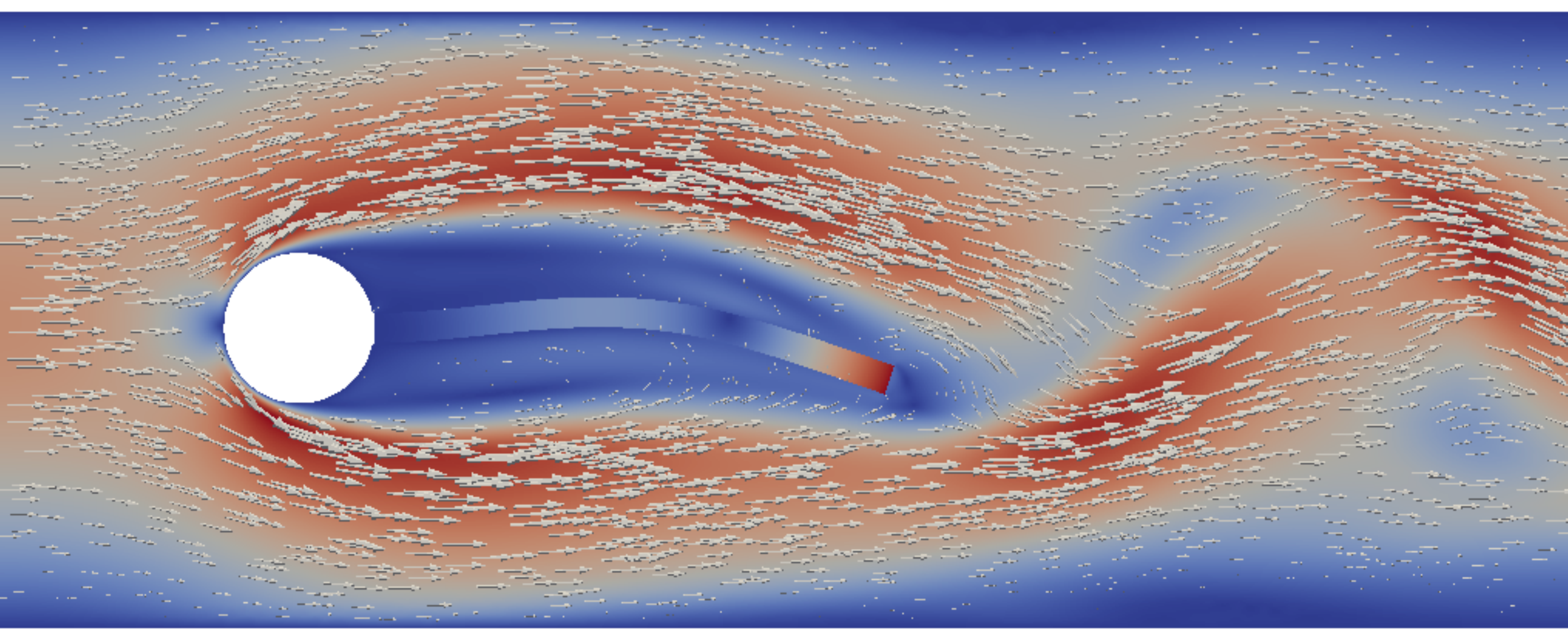}
	\caption{Method Validation. Portion of the domain proposed by \cite{Turek} in time $t=3\left[s\right]$. Deformation is represented on the bar and velocity field in the fluid domain. }
	\label{exposcillation}
\end{figure}

Validating the algorithm is a mandatory step to trust the results in this section. The numerical method is validated with the benchmark \textbf{FSI3} proposed in section 4.3 of \cite{Turek}. The experimental set-up involves a flexible rod oscillating in a fluid flow. brown The dimensions of the fluid domain are $41.0\times250.0[cm]$, the dimensions of the rod $2.0\times 35.0[cm]$ and the radius of the anchoring structure being $r=5.0[cm]$. The densities for the fluid and the solid are $\rho_f=\rho_s=1[g/cm^3]$. The dynamic viscosity for the fluid is $\mu_f=10.0[Poise]$ and the Young modulus and Poisson's ratio for the isotropic solid are $E=5.6E7 [Barye]$ and $\nu=0.4[-]$ respectively. The fluid and solid meshes are composed by $35k$ and $16k$ linear triangles respectively and the problem solved with a time step of $1E{-3}[s]$. Oscillation frequency and amplitude at the tip of the rod are measured to compare against numerical results.Results for t=3[s] are shown in \cref{exposcillation}.

For our code, the obtained amplitude and frequency on the quasi-periodic period are $A_x=-2.60\times10^{-3}\pm2.40\times10^{-3}\left[f=10.8\right]$ and $A_y=2.3\times10^{-3}\pm33.7\times10^{-3}\left[f=5.4\right]$  in concordance with the $A_x=-2.69\times10^{-3}\pm2.53\times10^{-3}\left[f=10.9\right]$ and $A_y=1.48\times10^{-3}\pm34.38\times10^{-3}\left[f=5.3\right]$ obtained in the cited experiment\footnote{The results are presented as in the original experiment: mean $\pm$amplitude$\left[freq\right]$.}. Although there is already a good agreement, results can be further improved by refining the meshes, as proven in section 4.1 of \cite{Cajas2015}. With this, we prove the algorithm is correctly implemented and reproduces the physics of the FSI problem.
%

\subsection{Wave propagation in elastic tube}
\label{subsec:exp:tube}
\begin{figure}[H]
	\centering
	\includegraphics[width=0.75\textwidth]{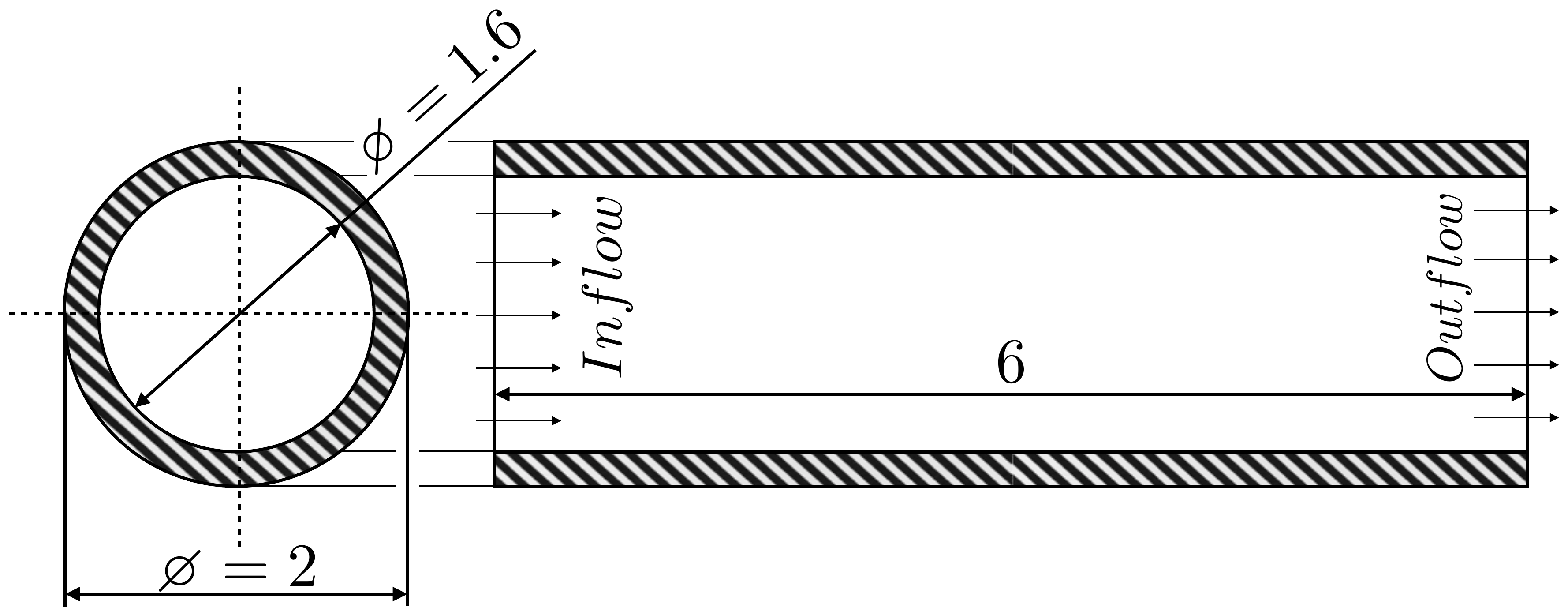}
	\caption{ Scheme with dimensions for the wave propagation in elastic tube experiment. \label{fig:res:tube_scheme}}
\end{figure}

The domain, schematised in \cref{fig:res:tube_scheme}, is an elastic tube, filled with fluid. The densities are $\rho_f=\rho_s=1[g/cm^3]$ for the fluid and the solid. Fluid viscosity is $\mu=0.03[Poise]$. The Young modulus and Poisson's ratio for the solid are $E=3E7[Baryes]$ and $\nu=0.3[-]$. Inflow velocity is $30[cm/s]$. Outflow pressure is $p=0[Baryes]$. In the contact surface, continuity of displacement and normal stresses are imposed. Linear tetrahedra are used for the spatial discretisation resulting in 48k elements (10k nodes) and 30k elements (6k nodes) for the fluid and the solid respectively, with 2.7k interface nodes ($\sim$25\% of the total). Time step is fixed at $\Delta t=$4E-4$[s]$. For each case a sensitivity analysis is done with a range of previous time steps, iterations and $ \epsilon$.  Each case run in 24 cores in Marenostrum IV. As a reference for the reader,  with the optimal configurations the CIQN algorithm case took 38 minutes and the Aitken case took 67 minutes 13 seconds.

\begin{figure} [H]
	\centering
	\begin{subfigure}[b]{0.475\textwidth}
		\centering			
		\includegraphics[width=\textwidth]{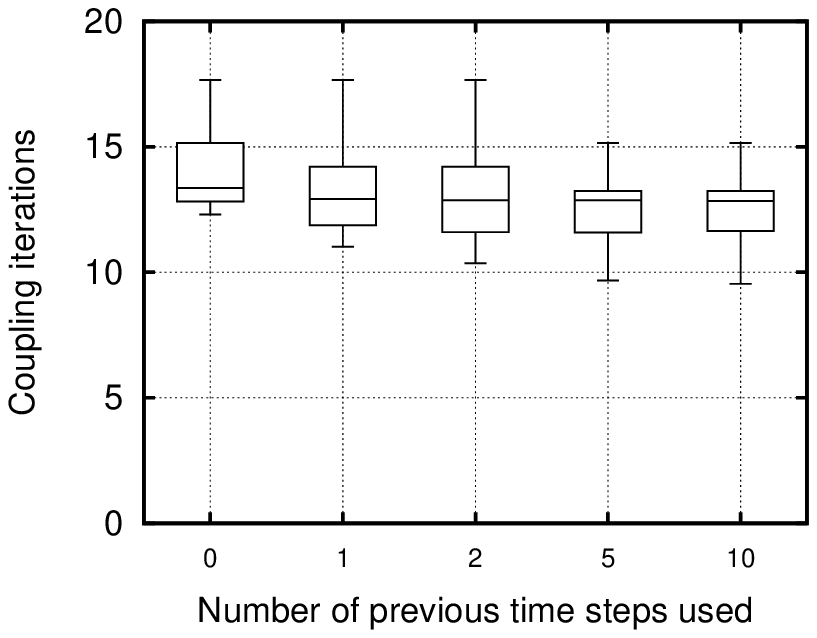}
	\end{subfigure}
	\hfill
	\begin{subfigure}[b]{0.475\textwidth}  
		\centering 
		\includegraphics[width=\textwidth]{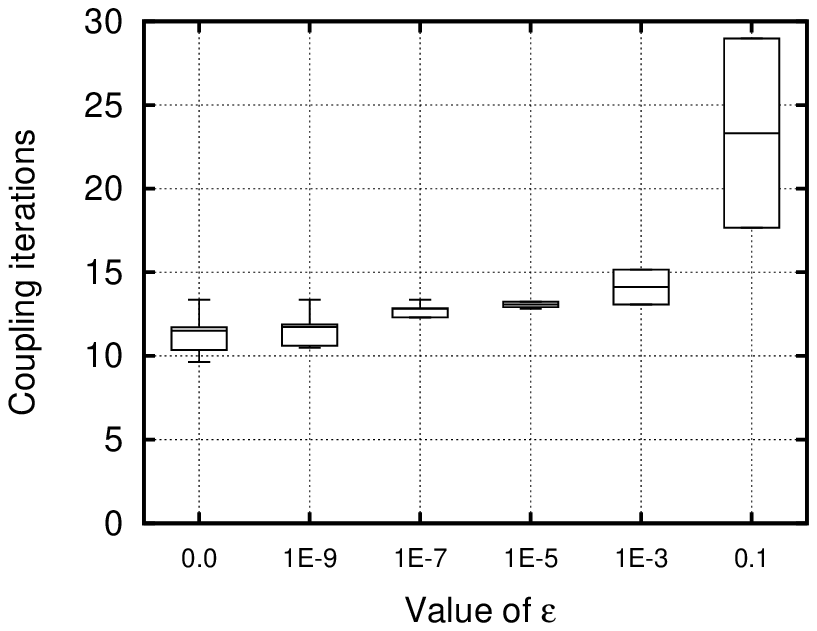}
	\end{subfigure}
	\caption{Candle plots for the wave propagation in an elastic tube experiment when displacement is relaxed. Less coupling iterations are required with a larger number of previous time steps and a smaller $\epsilon$.\label{fig:candles_tube}} 
	
\end{figure}

Results and statistical tendencies relaxing displacement are shown in \cref{table:tube} and \cref{fig:candles_tube}. Similar information, but relaxing force is shown in \ref{subsec:appx:tube}. For a qualitative comparison, with the Aitken  algorithm, the solver requires 17.76 (sd=2.91) and 21.81 (sd=3.64) iterations in average when relaxed on force and displacement respectively.

\begin{table}[H]
	
	\begin{center}
		\begin{tabular}{c|c||cccccc}
			
		histories	& ranking & $\epsilon$=0&$\epsilon$=1E-9&$\epsilon$=1E-7&$\epsilon$=1E-5&$\epsilon$=1E-3&$\epsilon$=0.1\\
			\hline
			\rowcolor{LightGray}
			      & 5 &  13.36 & 13.36 & 13.36 & 13.24 & 13.08 & 17.66\\
			 \rowcolor{LightGray}
			     \multirow{-2}{*}{0} & 10 & 12.32 &  12.32 &  12.30 &  12.82 &  15.16 &  28.98 \\
			 %
			 %
			      & 5 & 11.25 & 11.02 &  12.86 & 13.24 & 13.08  & 17.66\\
			     \multirow{-2}{*}{1} & 10 & 11.72 & 11.88 &  12.41 &  12.92 &  15.16 &  28.97 \\
			 %
			 \rowcolor{LightGray}
			       & 5 &  10.36 & 10.55 & 12.82 &  13.24 & 13.08 & 17.66 \\
			 \rowcolor{LightGray}
			      \multirow{-2}{*}{2} & 10 & 11.43  &  11.57 & 11.77 & 12.92 & 15.16 & 28.97 \\
			 %
			 %
			       & 5 &  9.67 & 10.62 &  12.87 &  13.24 & 13.08 & 17.66 \\
			     \multirow{-2}{*}{5} & 10 & 11.59 &  11.82 &  12.31 &  12.89 &  15.16 &  28.97 \\
			 %
			 %
			 \rowcolor{LightGray}
			      & 5 & 9.64 & 10.49 & 12.84 & 13.24 & 13.08 & 17.66\\
			 \rowcolor{LightGray}
			     \multirow{-2}{*}{10} & 10 &  11.64 & 11.68 &  12.31 & 12.92 &  15.16 &  28.97 \\

			\hline
			
		\end{tabular}
	\end{center}
	\caption{Results for the wave propagation in an elastic tube experiment when displacement is relaxed. Iterations for the scheme depending on the number of previous time steps used (histories), iterations in each time step (ranking) and filter ($\epsilon$) when relaxing displacement. The average number of iterations is 14.2925 (sd=4.72). \label{table:tube}}
\end{table}

Although simple, the problem in this section has been similarly reproduced in other FSI articles \cite{Causin2005, Degroote2008, Degroote2013, Haelterman2016}. Here we show that, as similarly concluded in \cite{Haelterman2016} adding information from previous time steps improves the rate of convergence of the algorithm (left plot on \cref{fig:candles_tube}). On the contrary, increasing the ranking does not necessarily have a positive effect. Filtering has the effect of reducing the standard deviation on the number of iteration in each time step  (right plot on \cref{fig:candles_tube}), but it is arguable if it compensates the added computational cost of restarting the QR decomposition. A similar behaviour can be seen if forced is relaxed (see \ref{subsec:appx:tube}).

\subsection{Oscillating rod and flexible wall in a fluid domain}
\label{subsec:exp:angrybird}

\begin{figure}[H]
	\centering
	\includegraphics[width=0.45\textwidth]{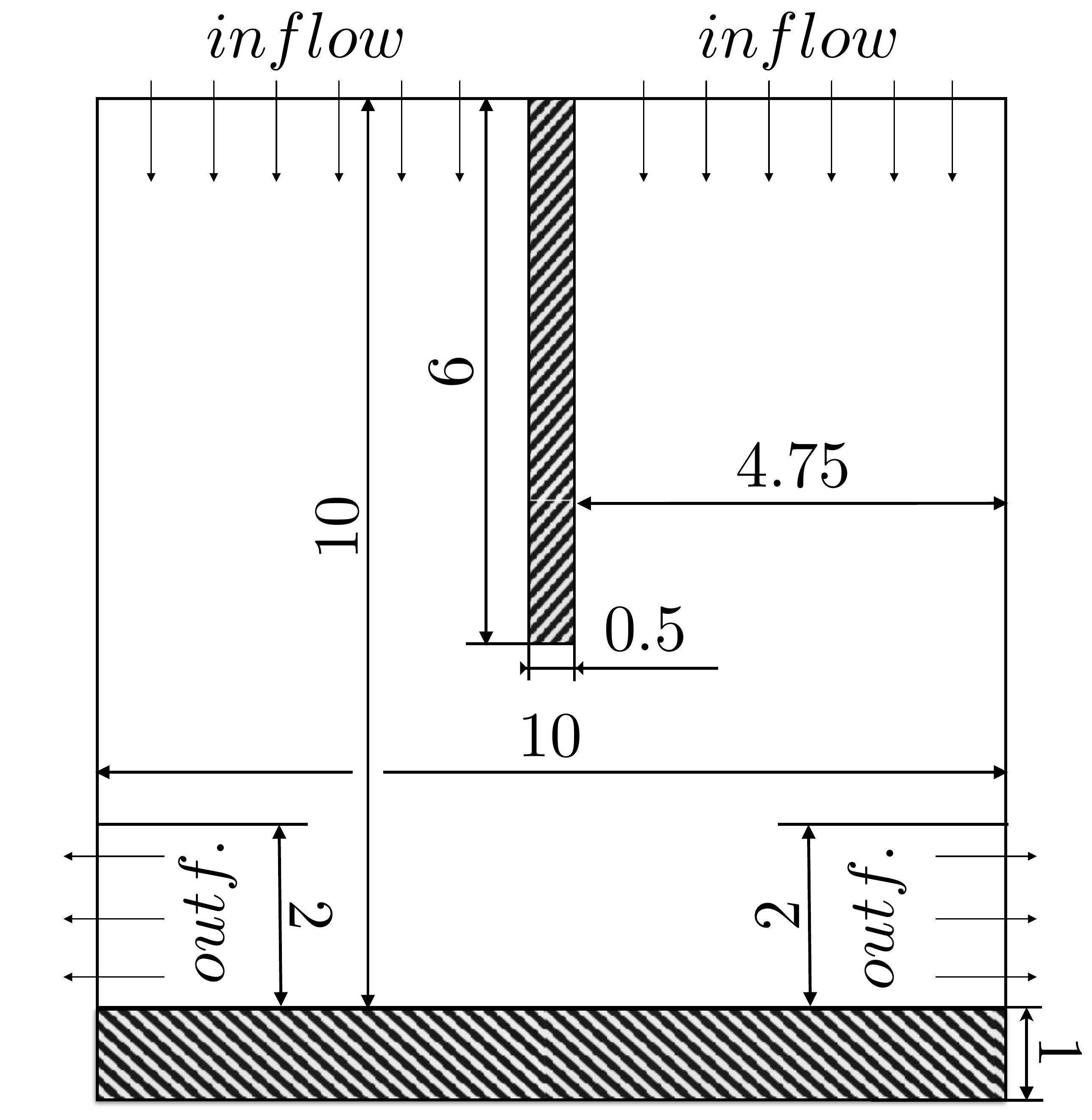}
	\caption{Scheme with dimensions for the oscillating rod with flexible wall experiment. \label{fig:res:ab_scheme}}
\end{figure}

The domain, schematised in \cref{fig:res:ab_scheme}, is composed by a centered oscillating flexible rod and a fixed flexible wall, being the rest the fluid domain. The densities for the fluid and the solid are $\rho_f=\rho_s=1[g/cm^3]$ . Fluid viscosity is $\mu=0.04[Poise]$. The Young modulus and poisson ratio for the solid are $E=2E7[Baryes]$ and $\nu=0.3[-]$. The tip of the oscillating rod has an imposed dispalcement of $d_x=sin(2\pi t)$. The Inflows velocity is $0.1[cm/s]$. Outflows pressures are $p=0[Baryes]$. Continuity of displacement and normal stresses are imposed in both contact surfaces. Linear triangles are used for the spatial discretisation resulting in 7.4k elements (4k nodes) and 752 elements (474 nodes) for the fluid and the solid respectively, with 300 interface nodes ($\sim$10\% and $\sim$45\% of the total for the fluid and the solid respectively). Time step is fixed at $\Delta t=0.1[s]$. Each case run in 16 cores in Marenostrum IV. As a reference for the reader, with the optimal configurations, the CIQN algorithm case took 57 seconds and the Aitken case took 7 minutes and 49 seconds.

\begin{figure}[H]
	\centering
	\begin{subfigure}[b]{0.475\textwidth}
		\centering			
		\includegraphics[width=\textwidth]{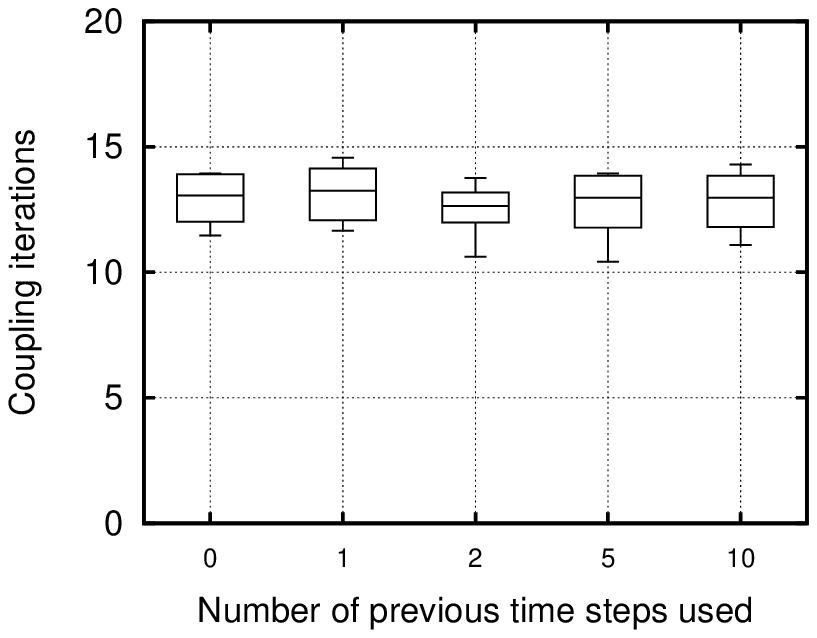}
	\end{subfigure}
	\hfill
	\begin{subfigure}[b]{0.475\textwidth}  
		\centering 
		\includegraphics[width=\textwidth]{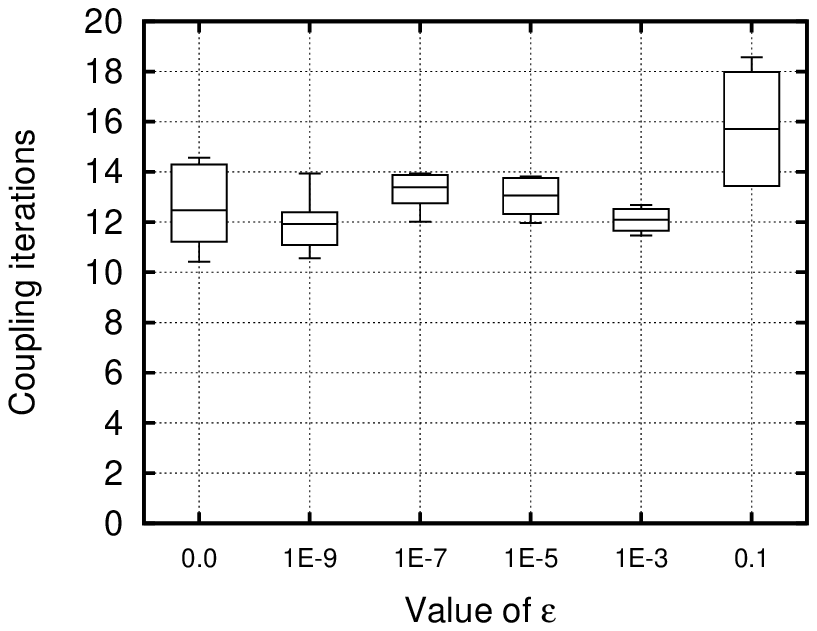}
	\end{subfigure}
	\caption{Candle plots for the oscillating rod experiment when displacement is relaxed. The number of coupling iterations do not seem to improve with filtering or re-usage of previous information.\label{fig:candles_ab}} 
	
\end{figure}

Results and statistical tendencies relaxing displacement are shown in \cref{table:ab} and \cref{fig:candles_ab}. Similar information, but relaxing force is shown in \ref{subsec:appx:angrybird}. For a qualitative comparison, with the Aitken algorithm, the solver requires  55.46 (sd=41.43) and  69.98 (sd=55.94) iterations in average when relaxed on force and displacement respectively.

\begin{table}[H]
	
	\begin{center}
		\begin{tabular}{c|c||cccccc}
			histories	& ranking & $\epsilon$=0&$\epsilon$=1E-9&$\epsilon$=1E-7&$\epsilon$=1E-5&$\epsilon$=1E-3&$\epsilon$=0.1\\
			\hline
			\rowcolor{LightGray}
             & 5 & 12.02 & 12.02 & 12.02 & 11.96 & F & 13.44 \\
            \rowcolor{LightGray}
            \multirow{-2}{*}{0} & 10 & 13.94 & 13.94 & 13.88 & 13.82 & 12.68 & 17.98 \\
             & 5 & 14.56 & 14.52 & 13.14 & 12.36 & 11.66 & 13.44 \\
            \multirow{-2}{*}{1} & 10 &11.8 & 11.74 & F & 13.76 & 12.52 & 17.98 \\
            \rowcolor{LightGray}
             & 5 & 12.92 & 12.4 & 12.76 & 12.32 & 11.66 & 13.44 \\
            \rowcolor{LightGray}
            \multirow{-2}{*}{2} & 10 & 10.62 & 10.92 & 12.76 & 13.76 & 12.52 & 17.98\\
             & 5 & F & 11.9 & 13.42 &  12.32 & 11.66 & 13.44\\
            \multirow{-2}{*}{5} & 10 & 10.42 & 10.56 & 13.94 & 13.76 & 12.52 & F\\
            \rowcolor{LightGray}
             & 5 & 14.30 & 11.96 & 13.42 & 12.32 & 11.66 & 13.44\\
            \rowcolor{LightGray}
            \multirow{-2}{*}{10} & 10 & 12.22 & 12.34 & 13.94 & 13.76 & 12.52 & 17.98\\
			\hline
			
		\end{tabular}
	\end{center}
	\caption{ Results for the oscillating rod experiment when displacement is relaxed. Iterations for the scheme depending on the number of previous time steps used (histories), iterations in each time step (ranking) and filter ($\epsilon$) when relaxing displacement, F meaning a diverged simulation. The average number of iterations is 14.99 (sd=8.33). \label{table:ab}}
\end{table}

The problem presented in this section, even though being computationally cheaper, is more physically challenging. Compared to problem in \cref{subsec:exp:tube}, there are two surfaces to couple and the deformations are considerably larger. This is reflected in the computational cost, requiring 55.46 (sd=41.43) Aitken iterations \footnote{compared to the 17.76 (sd=2.91) Aitken iterations required by the elastic tube experiment.}. In this case, the behaviour of the IQN algorithm is completely different from the presented in \cref{subsec:exp:tube}. Now we can see scattered diverging simulations in \cref{table:ab}, increasing the number of histories do not bring any advantage (left side of \cref{fig:candles_ab}) and using filtering doesn't bring any drastic improvement (right side of \cref{fig:candles_ab}). On the contrary, and oppositely to the case presented in \cref{subsec:exp:tube}, increasing the rank has a beneficial effect on the convergence properties of the algorithm when little or no filtering is included. The strongest hypothesis for this behaviour is that including histories over-predicts the final position of the interface, hampering the convergence of the algorithm.

\subsection{Scalability}
\label{subsec:exp:scalability}

\begin{figure}[H]
	\centering
	\includegraphics[width=1.0\textwidth]{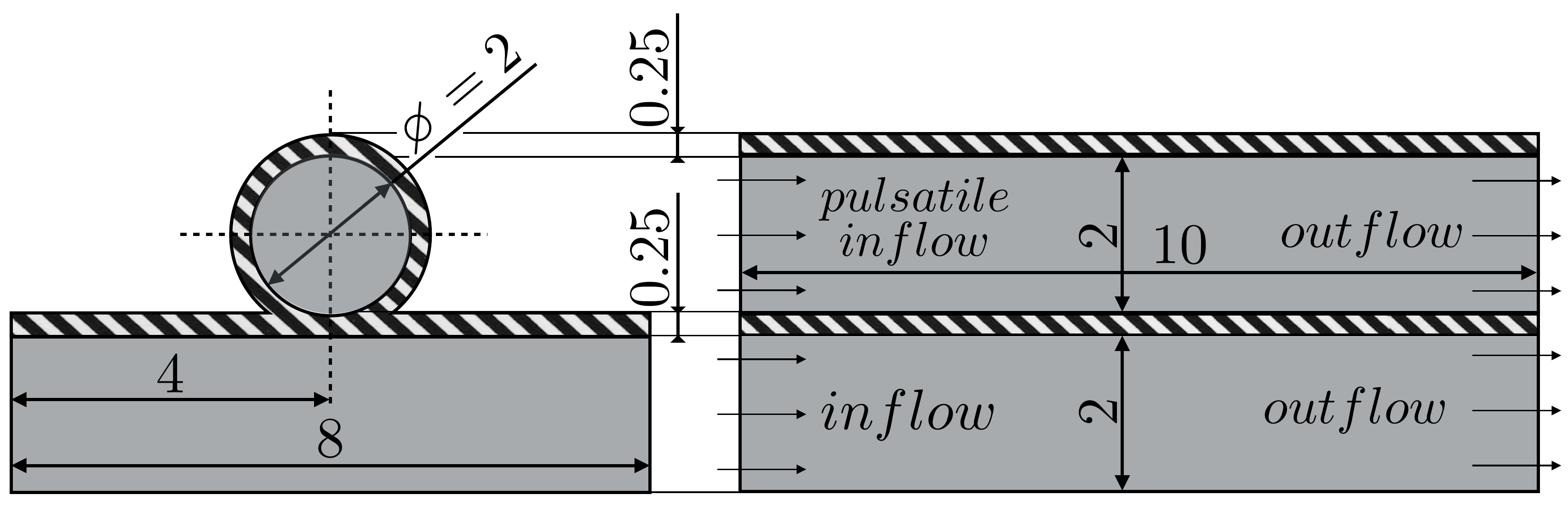}
	\caption{Scheme with dimensions for the scalability case. \label{fig:res:art_scheme}}
\end{figure}

The domain, schematised in \cref{fig:res:art_scheme}, is a filled flexible tube lying over a flexible surface which is in contact with a big volume of another fluid.  The densities for the fluid and the solid are $\rho_f=\rho_s=1[g/cm^3]$. Fluid viscosity is $\mu=0.03[Poise]$. The Young modulus and Poisson ratio for the solid are $E=1.5E4[Baryes]$ and $\nu=0.3[-]$. The Inflow velocites are $sin(2\pi t)$ and $1[cm/s]$ for the lower domain respectively. Outflows pressures are $p=0[Baryes]$. Continuity of displacement and normal stresses are imposed in both contact surfaces. Linear tetrahedra are used for the spatial discretisation resulting in  60M elements (10.4M nodes) and 40M elements (7.1M nodes) for the fluid and the solid respectively, with 4M interface nodes ($\sim$38\% and  $\sim$56\% of the total nodes for the fluid and the solid respectively). Time step is fixed at $\Delta t=0.1[s]$.

\begin{figure}[H]
	\centering
	\includegraphics[width=0.7\textwidth]{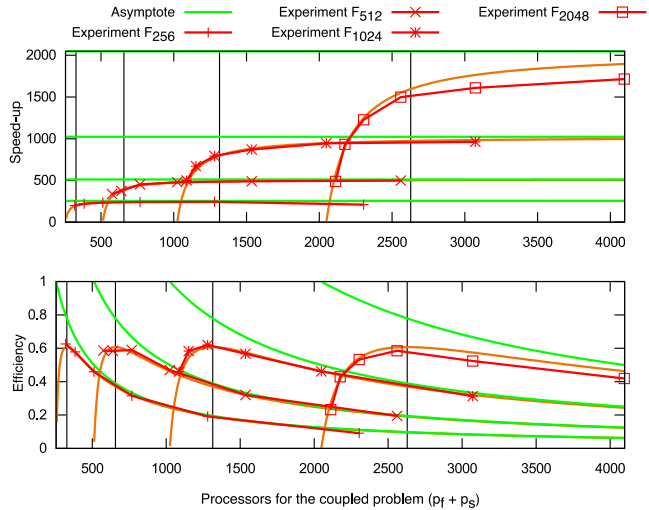}
	\caption{Parallel performance analysis. Speed-up and Efficiency for a core allocation of $p=p_f+p_s$, where $p_f = \{256$, $512$, $1024$, ${2048}\}$ in the core count for the fluid, and $p_s = \{64$, $128$, $256$, $512$, ${1024}$, ${2048}\}$ for the solid. The orange line is the fitted curve for each case. An optimal allocation $p_{opt}$ which allows to achieve the maximum efficiency $E_{p_f}$ of the coupled system can be found for each curve $S_{p_f}$. }
	\label{fig:scala_coupeld}
\end{figure}

As the main goal is to prove good scalability, the results for the sensibility analysis are shown in \ref{subsec:appx:scalability}. \Cref{table:scala} show results for the solver running independently (uncoupled) to ease its comparison with the coupled scalability. To obtain an optimum efficiency $E$, several cases are run sweeping the core allocation for each physical problem. \Cref{fig:scala_coupeld} shows speed-up and efficiency for four fixed values in the fluid solver core count $p_f$. In each case the core count for the solid mechanic solver $p_s$ is ranged between $64$ and $2048$, with increments in power of two. This processes is performed for a core count of $256$, $512$, $1024$, and $2048$ in the fluid solver. The results is a set of curves with a peak efficiency $E_{p_f}$ given by the optimal balance of cores for each case.

\begin{table}[H]
	\begin{center}
		\begin{tabular}{c|cc|cc}
			
			Core	& \multicolumn{2}{c|}{Fluid mechanics}& \multicolumn{2}{c}{Solid mechanics} \\
			count	&  speed up & efficiency &speed up &efficiency \\
			\hline
			\rowcolor{LightGray}
		     128     & 128.0  		& 1.00 & 128.0  & 1.00 \\
   		     256     & 256.0 		& 0.99 &  256.0 & 0.99 \\
   		     \rowcolor{LightGray}
   		     512     & 511.6        & 0.99 &  508.1 & 0.99 \\
  		     1024    & 1011.0      & 0.98 & 960.5 & 0.93\\
  		     \rowcolor{LightGray}
  		     2048    & 1880.3      & 0.91 & 1793.0 & 0.87\\			
			\hline
		\end{tabular}
	\end{center}
	\caption{Parallel performance analysis. Speed-up and efficiency for both solvers.\label{table:scala}}
\end{table}

This example demonstrates the necessity of the parallel version of the algorithm, as it would be impossible to fit 11M interface nodes in a single node of a shared-memory high-performance computing infrastructure. Furthermore, we show a scalability up to 4800 cores. Although the scalability of the uncoupled problem do not drop under 87\% in the uncoupled case, when coupled, the maximum scalability achieved is 60\%. This is due to the staggered scheme used that improves stability, one set of cores is idle while the rest is processing.  Although it's out of the scope of this article, this issue can be tackled by using core overloading on the MPI scheme.

For another large scale use of the algorithm, please refer to \cite{Santiago2018}, where a human heart is solved with the IQN algorithm.



\section{Conclusions \& future work}
\label{sec:conclusion_and_fw}

In this paper, we introduced the compact interface quasi-Newton (CIQN) coupling scheme, optimised for distributed memory architecture. The developed algorithm includes reusing information from the previous time steps (histories) and filtering, but in an efficient scheme that avoids constructing dense matrices and reduces the number of operations. This leads to an algorithm that requires less coupling iterations and computing time per time step.

In previous works \cite{uekermann2016partitioned,Haelterman2016} it has been stated that using information from previous time steps together with filtering improve convergence. In this work we show this do not necessarily happen and can only be stated for certain type of dynamic behaviour, while a correct parametrisation requires a fine tuning for each case.

In this work we prove that a parallel coupling algorithm is mandatory for massively parallel cases. We also show that reusing histories does not necessarily improves the convergence rate of the algorithm, but is dependant on the dynamics of the problem. As it has been said, there is no a silver bullet algorithm to tackle all the FSI cases. The chosen algorithm and parameters must fit the features of the problem to solve, taking into account the dynamics and the possible numerical instabilities that may arise.

Although the developed algorithm has been proved robust and efficient, there is room for optimise the execution with core-overloading. Also, other filtering algorithms (\textit{e.g.} eigenvalue-based) can be tested to compare performance and computational cost. Finally, the behaviour of the presented algorithm has to be tested in other interface problems like solid-solid contact or heat transmission. These topics will be developed in a future work.


\section*{Acknowledgements}
This work has been funded by CompBioMed project a grant by the European Commission H2020 (agreement nr: 823712), EUBrazilCC a project under the Programme FP7-ICT (agreement nr: 614048) and a FPI-SO grant (agreement nr: SVP-2014-068491).

\appendix

\section{Glossary}
\label{sec:appx:glossary}
Glossary of acronyms used at the manuscript.
\begin{itemize}
    \item \textbf{ALE:} Arbitrary Lagrangian-Eulerian.
    \item \textbf{BSC:} Barcelona Supercomputing Center.
    \item \textbf{CFD:} Computed Fluid Dynamics.
    \item \textbf{CIQN:} Compact interface quasi-Newton.
    \item \textbf{CSM:} Computed Solid Mechanics.
    \item \textbf{$E$:} Young modulus.
    \item \textbf{FSI:} Fluid-structure interaction.
    \item \textbf{HPC:} High Performance Computing.
    \item \textbf{IQN:} Interface quasi-Newton.
    \item \textbf{MPI:} Message Passing Interface.
    \item \textbf{sd:} Standard deviation.
    \item \textbf{$\nu$:} Poisson's ratio.
    \item \textbf{$\rho$:} Density.
    \item \textbf{$\mu$:} Dynamic viscosity.
\end{itemize}

\section{Index notation convention}
\label{sec:appx:index_convention}
To ease implementation, the Einstein convention on repeated indices will be followed, allowing to describe the mathematics, the physics and the computational implementation aspects depending on the context. For the continuum problem, the indices label space dimensions. On the discretised problem, the lowercase greek alphabet $\alpha =1,\cdots p$ labels the number of degrees of freedom $p$, i.e. the matrix rows. The lowercase latin alphabet labels the matrix columns, $i=1,\cdots q-1$ where $q$ is the last stored iteration. Additionally, a capital latin subindex labels the FSI solver iteration $I=1,\cdots q-1$, where $q$ is the last stored iteration. A final rule is how those indices operate: only those of the same kind are contracted. For instance, $Q^{I-1}_{\alpha i}$ is the $Q$ matrix for iteration $I-1$ with rows labelled $\alpha$ and columns $i$. When this matrix is multiplied by a certain vector $B_{i}$, it results in a given vector $A^{I-1}_{\alpha}$:
\begin{equation}
	A^{I-1}_{\alpha}= Q^{I-1}_{\alpha i} B_{i} = \sum_{i=1}^{q-1} Q^{I-1}_{\alpha i} B_{i},\nonumber
\end{equation}
where latin indices $i$ are contracted.

\section{More results}
\label{sec:appx:results}

\subsection{Typical behaviour of the coupling residue and solver iterations}
\label{subsec:appx:typic}

\begin{figure}[H]
	\centering
	\includegraphics[width=0.8\textwidth]{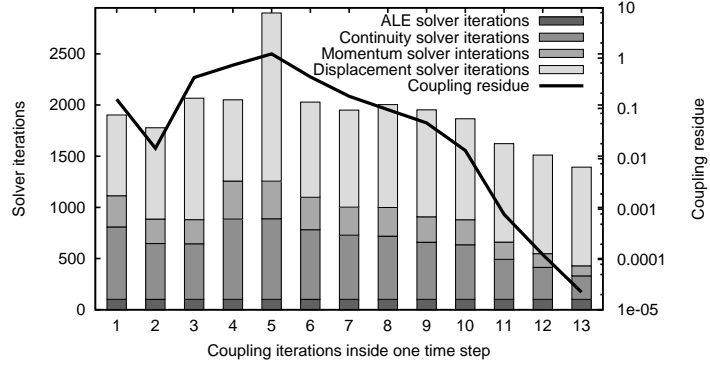}
	\caption{Iterations of the different solvers and resiude of the coupling for a typical time step in the CIQN algorithm. \label{fig:res:typic}}
\end{figure}

In \Cref{fig:res:typic} we show the typical behaviour of the coupling residue and solver iterations for a single time step. Particularly, it corresponds to the 20th time step of the case presented in \cref{subsec:exp:angrybird} using 5 iterations per time step (ranking=5) and 5 previous time iterations (history=5) and an $\epsilon$=1E-9.

%

\subsection{For the wave propagation in elastic tube}
\label{subsec:appx:tube}

\begin{figure}[H]
	\centering
	\begin{subfigure}[b]{0.475\textwidth}
		\centering			
		\includegraphics[width=\textwidth]{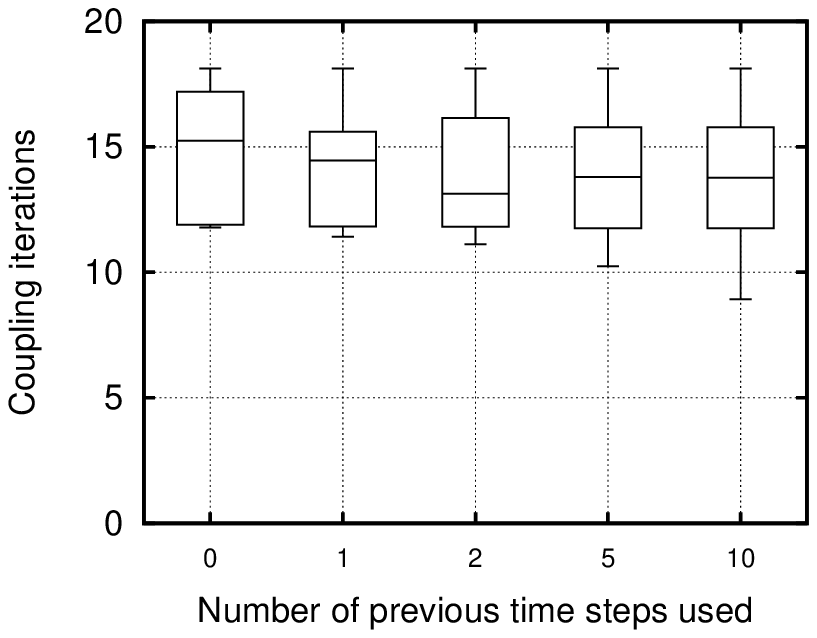}
	\end{subfigure}
	\hfill
	\begin{subfigure}[b]{0.475\textwidth}  
		\centering 
		\includegraphics[width=\textwidth]{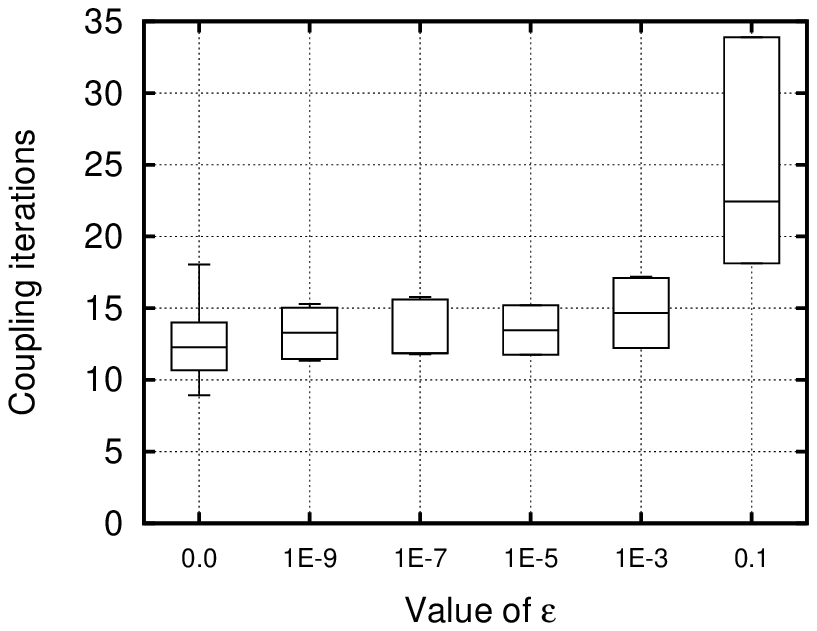}
	\end{subfigure}
	\caption{Candle plots for the wave propagation experiment when force is relaxed.\label{fig:candles_tubeF}} 
	
\end{figure}

 \Cref{table:tubeF,fig:candles_tubeF} show the results when force is relaxed.

\begin{table}[H]
	\begin{center}
		\begin{tabular}{c|c||cccccc}
			
			histories & ranking &$\epsilon$=0&$\epsilon$=1E-9&$\epsilon$=1E-7&$\epsilon$=1E-5&$\epsilon$=1E-3&$\epsilon$=0.1\\
			\hline
			\rowcolor{LightGray}
              & 5 & 11.89 & 11.84 &  11.79 &  11.80 &  12.22 &  18.12 \\
             \rowcolor{LightGray}
             \multirow{-2}{*}{0} & 10 &  15.24 & 15.29 &  15.55 &  15.13 & 17.19 & 33.90\\
              & 5 &  11.42 &    11.54 &    11.83 &   11.76 &   12.22 &   18.12 \\
             \multirow{-2}{*}{1} & 10 &   14.46 &     15.07 &    15.60 &     15.18 &    17.11 &    33.9\\
             \rowcolor{LightGray}
              & 5 &   11.33  &   11.87 &    11.76 &   11.12 &   12.22 &    18.12 \\
             \rowcolor{LightGray}
             \multirow{-2}{*}{2} & 10 &  13.99 &   15.04 &    15.78 &   15.20 &  17.11 &   33.90\\
              & 5 &  11.87 &  10.68 &   11.34 &   11.76 &   12.22 &   18.12\\
             \multirow{-2}{*}{5} & 10 &  13.80 &   14.89 &   15.78 &  15.20 & 17.11 &   26.74\\
             \rowcolor{LightGray}
              & 5 & 10.59 &   11.45 &   11.87 &   11.76 &   12.22 &   18.12 \\
             \rowcolor{LightGray}
             \multirow{-2}{*}{10} & 10 & 13.77 &   14.73 &   15.78 &  15.20 &  17.11 &  33.90\\
			\hline

		\end{tabular}
	\end{center}
	\caption{Results for the wave propagation in an elastic tube experiment when relaxing force. Iterations for the scheme depending on the number of previous time steps used (histories), iterations in each time step (ranking) and filter ($\epsilon$) when relaxing displacement.caption relaxing force. The average number of iterations is 15.298125(sd=5.58) \label{table:tubeF}}
\end{table}

%

\subsection{For the oscillating bar and flexible wall in a fluid domain}
\label{subsec:appx:angrybird}

\begin{figure}[H]
	\centering
	\begin{subfigure}[b]{0.475\textwidth}
		\centering			
		\includegraphics[width=\textwidth]{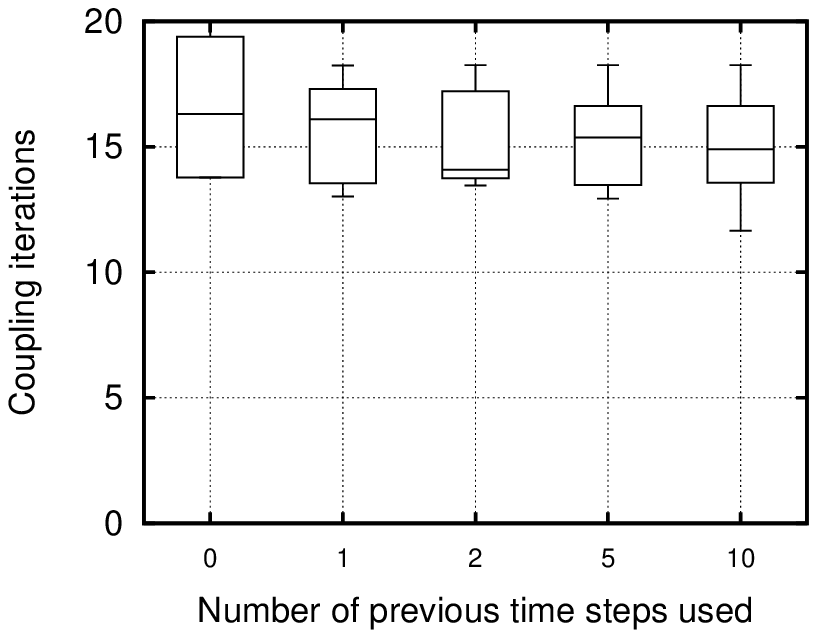}
	\end{subfigure}
	\hfill
	\begin{subfigure}[b]{0.475\textwidth}  
		\centering 
		\includegraphics[width=\textwidth]{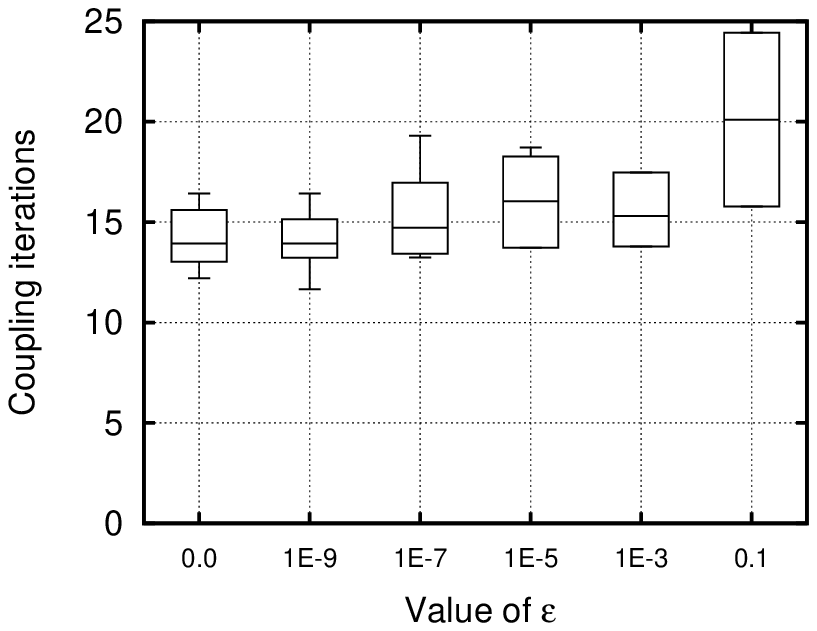}
	\end{subfigure}
	\caption{Candle plots for the oscillating rod experiment when force is relaxed. \label{fig:candles_abF}}
	
\end{figure}

 \Cref{table:abF,fig:candles_abF} show the results when force is relaxed.

\begin{table}[H]
	\begin{center}
		\begin{tabular}{c|c||cccccc}
			histories	& ranking &$\epsilon$=0&$\epsilon$=1E-9&$\epsilon$=1E-7&$\epsilon$=1E-5&$\epsilon$=1E-3&$\epsilon$=0.1\\
			\hline
			\rowcolor{LightGray}
              & 5 & 13.78 &  13.78  &  13.78  &  13.84 &  13.78 &  15.78\\
             \rowcolor{LightGray}
             \multirow{-2}{*}{0} & 10 &  19.48 &  19.48 &  19.30 &  18.72 &  16.84 &  24.42\\
               & 5 & 13.02  &   13.22 &    13.38 &   13.72 &  13.78 &  15.78\\
              \multirow{-2}{*}{1} & 10 & 16.42 & 16.42 & 17.14 & 18.24 & 17.48 &  24.44 \\
             \rowcolor{LightGray}
              & 5 & 13.46 &  13.46 &  13.82 &  13.72 &  13.78 &  15.78 \\
             \rowcolor{LightGray}
             \multirow{-2}{*}{2} & 10  & F & F &   16.96 &  18.26 &   17.48 &   24.44\\
              & 5 &12.94 & 13.10 &  13.24 &  13.72 & 13.78 & 15.78\\
             \multirow{-2}{*}{5} & 10 & 15.60 &  15.14 & 15.62 & 18.26 & 17.48 & 24.44\\
             \rowcolor{LightGray}
              & 5 & 12.20 & 11.66 & 13.42 & 13.72 & 13.78 & 15.78 \\
             \rowcolor{LightGray}
             \multirow{-2}{*}{10} & 10 &  15.28 &   14.91 &   15.62 &   18.26  &  17.48 &  24.44\\
			\hline
			
		\end{tabular}
	\end{center}
	\caption{Results for oscillating rod experiment when force is relaxed. Iterations for the scheme depending on the number of previous time steps used (histories), iterations in each time step (ranking) and filter ($\epsilon$) when relaxing displacement, F meaning a divergent case. The average number of iterations is 16.04 (sd=3.19). \label{table:abF}}
\end{table}

%

\subsection{For the scalability}
\label{subsec:appx:scalability}

\begin{figure}[H]
	\centering
	\begin{subfigure}[b]{0.475\textwidth}
		\centering			
		\includegraphics[width=\textwidth]{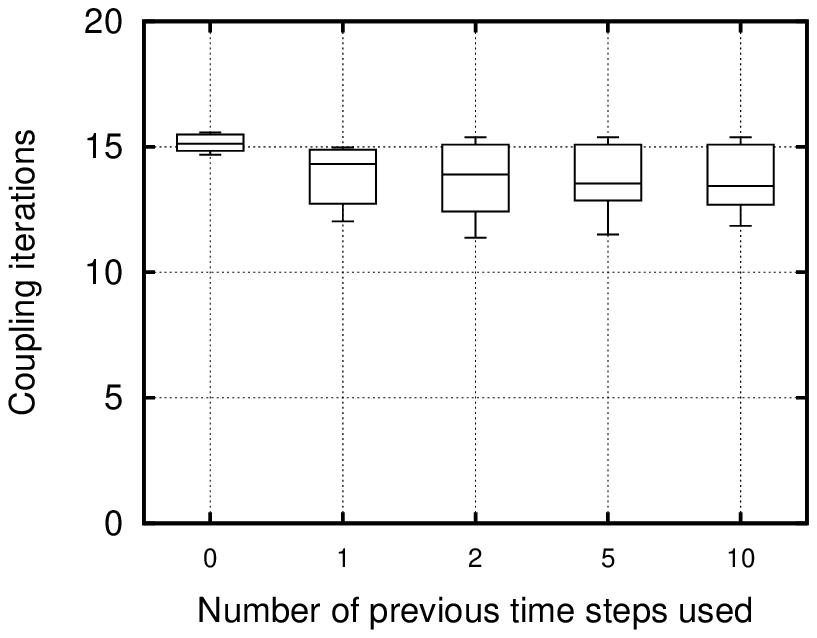}
	\end{subfigure}
	\hfill
	\begin{subfigure}[b]{0.475\textwidth}  
		\centering 
		\includegraphics[width=\textwidth]{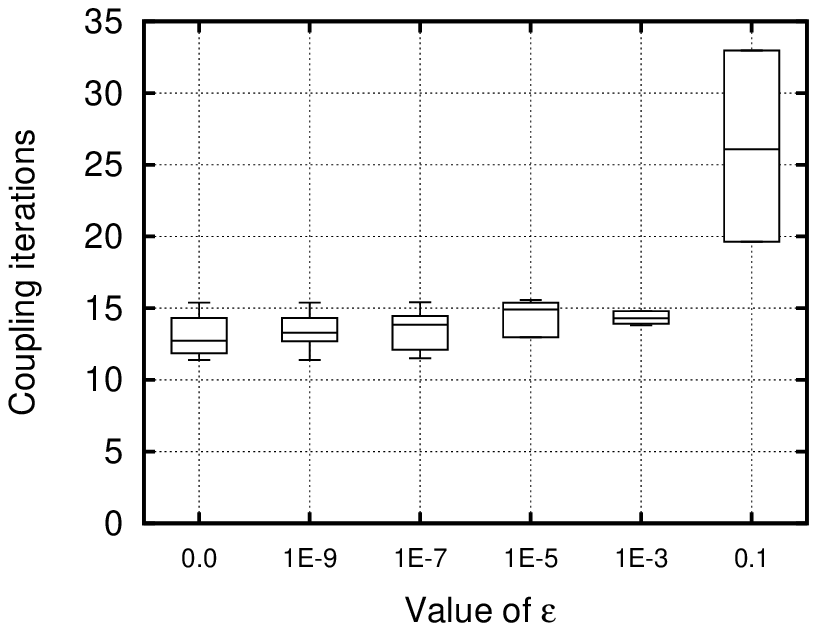}
	\end{subfigure}
	\caption{Candle plots for the scalability when displacement is relaxed. \label{fig:candles_art}}
	
\end{figure}

\begin{figure}[H]
	\centering
	\begin{subfigure}[b]{0.475\textwidth}
		\centering			
		\includegraphics[width=\textwidth]{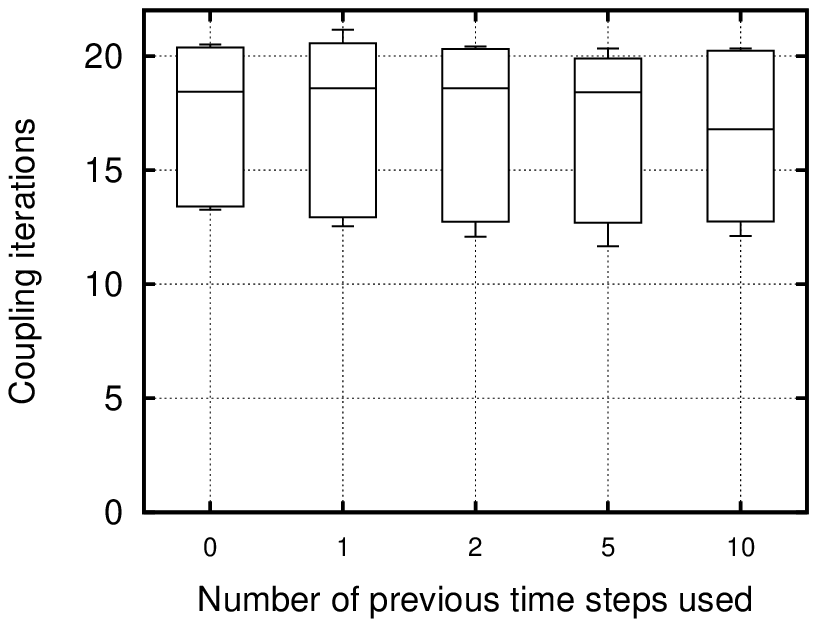}
	\end{subfigure}
	\hfill
	\begin{subfigure}[b]{0.475\textwidth}  
		\centering 
		\includegraphics[width=\textwidth]{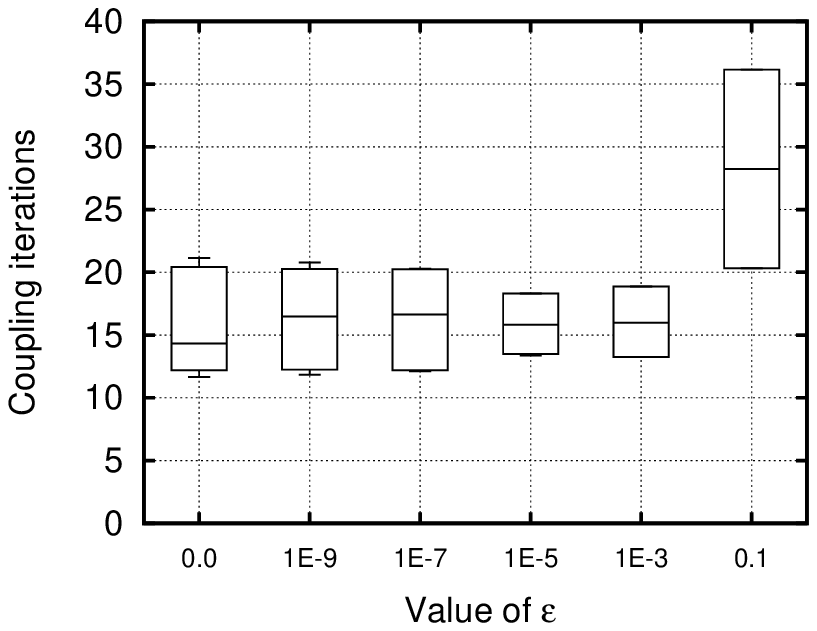}
	\end{subfigure}
	\caption{Candle plots for the scalability case when force is relaxed.\label{fig:candles_artF}}
	
\end{figure}

More results for the scalability case. As the objective of this experiment is to prove the HPC performance of the algorithm all the numerical sensitivity analysis is shown here. \Cref{fig:res:art_scheme} shows a scheme of the geometry used. \Cref{table:artery,fig:candles_art} show results when displacement is relaxed and \cref{table:arteryF,fig:candles_artF} when force is relaxed. For a quantitative comparison, with the Aitken algorithm the scheme requires 19.45 (sd=3.10) and 19.58 (sd=2.16) when relaxing on force and displacement respectively.  Each case run in 768 cores in Marenostrum IV.  As a reference for the reader, in the optimal configurations, the CIQN algorithm case took 1 hour, 27 minutes and 57 seconds and the Aitken case took 2 hours, 27 minutes and 30 seconds.

\begin{table}[H]
	
	\begin{center}
		\begin{tabular}{c|c||cccccc}
			histories	& ranking &$\epsilon$=0&$\epsilon$=1E-9&$\epsilon$=1E-7&$\epsilon$=1E-5&$\epsilon$=1E-3&$\epsilon$=0.1\\
			\hline
				\rowcolor{LightGray}
			  & 5 &  14.85 & 14.85 & 14.85 & 14.83 & 13.80 & 19.62\\
			\rowcolor{LightGray}
             \multirow{-2}{*}{0} & 10 & 15.39 & 15.39 & 15.41 & 15.57  & 14.69 & 32.56\\
              & 5 & 12.03  & 12.03 & 12.15  & 13.31  & 13.91  & 19.63\\
             \multirow{-2}{*}{1} & 10 & 14.31  & 14.31 & 14.46  & 14.98 & 14.79 & 32.97\\
			\rowcolor{LightGray}
              & 5 & 11.38 & 11.38  & 11.87 & 12.98 & 13.91  & 19.63 \\
			\rowcolor{LightGray}
             \multirow{-2}{*}{2} & 10 & 13.88 & 13.56  & 13.94 & 15.38  & 14.79 & 32.97\\
              & 5 & 11.81 & 13.23  & 11.51  & 12.97  & 13.91  & 19.63\\
             \multirow{-2}{*}{5} & 10 & 12.75 &  13.35 & 13.73  & 15.38  & 14.79 & 32.97\\
			\rowcolor{LightGray}
              & 5 &  11.85 & 12.69  & 12.11  & 12.97  & 13.91  & 19.63\\
			\rowcolor{LightGray}
             \multirow{-2}{*}{10} & 10 & 12.70 & 12.87  & 13.99  & 15.38 & 14.79  & 32.97\\
			
			\hline
			
		\end{tabular}
	\end{center}
	\caption{Results for scalability case when relaxing displacement. Iterations for the scheme depending on the number of previous time steps used (histories), iterations in each time step (ranking) and filter ($\epsilon$) when relaxing displacement. The average number of iterations is 15.80 (sd=5.52). \label{table:artery}}
\end{table}

\begin{table}[H]
	
	\begin{center}
		\begin{tabular}{c|c||cccccc}
			histories	& ranking &$\epsilon$=0&$\epsilon$=1E-9&$\epsilon$=1E-7&$\epsilon$=1E-5&$\epsilon$=1E-3&$\epsilon$=0.1\\
			\hline
				\rowcolor{LightGray}
             & 5 & 13.40 & 13.40 & 13.40 & 13.36 & 13.26  & 20.33\\
				\rowcolor{LightGray}
             \multirow{-2}{*}{0} & 10 & 20.50 & 20.42 & 19.95 & 18.17 & 18.69 & 36.15\\
              & 5 & 12.60 & 12.60 & 12.53 & 13.50 & 13.26 & 20.33\\
            \multirow{-2}{*}{1} & 10 & F & 20.78  & 19.89 & 18.30 & 18.87 & 36.15\\
				\rowcolor{LightGray}
              & 5 & 12.13  & 12.08 & 12.21  & 13.49  & 13.26 & 20.33\\
				\rowcolor{LightGray}
             \multirow{-2}{*}{2} & 10 & F & 20.25 & 20.28  & 18.30 & 18.87 & 36.15\\
              & 5 &  11.66 &  11.83 & 12.12  & 13.49 & 13.26 &  20.33\\
             \multirow{-2}{*}{5} & 10 & F & F & 20.24  & 18.30 & 18.87  & 36.15\\
				\rowcolor{LightGray}
              & 5 & 12.19  & 12.24 & 12.11 & 13.49 & 13.26 & 20.33\\
				\rowcolor{LightGray}
             \multirow{-2}{*}{10} & 10 & F  & 20.23 & 20.24  & 18.30 & 18.87 & 36.15\\

			\hline
			
		\end{tabular}
	\end{center}
	\caption{Results for scalability case when relaxing force. Iterations for the scheme depending on the number of previous time steps used (histories), iterations in each time step (ranking) and filter ($\epsilon$) when relaxing displacement. F meaning a divergent experiment. The  average number of iterations is 18.09 (sd=6.37). \label{table:arteryF}}
\end{table}

%


\bibliography{./sections/05_references.bib}

\begin{thebibliography}{32}
\expandafter\ifx\csname natexlab\endcsname\relax\def\natexlab#1{#1}\fi
\providecommand{\url}[1]{\texttt{#1}}
\providecommand{\href}[2]{#2}
\providecommand{\path}[1]{#1}
\providecommand{\DOIprefix}{doi:}
\providecommand{\ArXivprefix}{arXiv:}
\providecommand{\URLprefix}{URL: }
\providecommand{\Pubmedprefix}{pmid:}
\providecommand{\doi}[1]{\href{http://dx.doi.org/#1}{\path{#1}}}
\providecommand{\Pubmed}[1]{\href{pmid:#1}{\path{#1}}}
\providecommand{\bibinfo}[2]{#2}
\ifx\xfnm\relax \def\xfnm[#1]{\unskip,\space#1}\fi
\bibitem[{Deparis et~al.(2006)Deparis, Discacciati, Fourestey, and
  Quarteroni}]{Deparis2006}
\bibinfo{author}{S.~Deparis}, \bibinfo{author}{M.~Discacciati},
  \bibinfo{author}{G.~Fourestey}, \bibinfo{author}{A.~Quarteroni},
\newblock \bibinfo{title}{{Fluid-structure algorithms based on
  Steklov-Poincar{\'{e}} operators}},
\newblock \bibinfo{journal}{Computer Methods in Applied Mechanics and
  Engineering} \bibinfo{volume}{195} (\bibinfo{year}{2006})
  \bibinfo{pages}{5797--5812}.
\bibitem[{Gee et~al.(2010)Gee, Kuttler, and Wall}]{Gee2010}
\bibinfo{author}{M.~W. Gee}, \bibinfo{author}{U.~Kuttler},
  \bibinfo{author}{W.~A. Wall},
\newblock \bibinfo{title}{{Truly monolithic algebraicmultigrid for
  fluid–structure interaction}},
\newblock \bibinfo{journal}{International Journal for Numerical Methods in
  Biomedical Engineering} \bibinfo{volume}{85} (\bibinfo{year}{2010})
  \bibinfo{pages}{987--1016}.
\bibitem[{Crosetto et~al.(2011)Crosetto, Reymond, Deparis, Kontaxakis,
  Stergiopulos, and Quarteroni}]{Crosetto2011}
\bibinfo{author}{P.~Crosetto}, \bibinfo{author}{P.~Reymond},
  \bibinfo{author}{S.~Deparis}, \bibinfo{author}{D.~Kontaxakis},
  \bibinfo{author}{N.~Stergiopulos}, \bibinfo{author}{A.~Quarteroni},
\newblock \bibinfo{title}{{Fluid – structure interaction simulation of aortic
  blood flow}},
\newblock \bibinfo{journal}{Computers and Fluids} \bibinfo{volume}{43}
  (\bibinfo{year}{2011}) \bibinfo{pages}{46--57}.
\bibitem[{Hron and Turek(2006)}]{Hron2006a}
\bibinfo{author}{J.~Hron}, \bibinfo{author}{S.~Turek},
\newblock \bibinfo{title}{A monolithic fem/multigrid solver for an ale
  formulation of fluid-structure interaction with applications in
  biomechanics},
\newblock \bibinfo{journal}{Fluid-Structure Interaction} \bibinfo{volume}{53}
  (\bibinfo{year}{2006}) \bibinfo{pages}{146--170}.
\bibitem[{Degroote et~al.(2009)Degroote, Bathe, and Vierendeels}]{Degroote2009}
\bibinfo{author}{J.~Degroote}, \bibinfo{author}{K.-j. Bathe},
  \bibinfo{author}{J.~Vierendeels},
\newblock \bibinfo{title}{{Performance of a new partitioned procedure versus a
  monolithic procedure in fluid – structure interaction}},
\newblock \bibinfo{journal}{Computers and Structures} \bibinfo{volume}{87}
  (\bibinfo{year}{2009}) \bibinfo{pages}{793--801}.
\bibitem[{Matthies and Steindorf(2003)}]{Matthies2003}
\bibinfo{author}{H.~G. Matthies}, \bibinfo{author}{J.~Steindorf},
\newblock \bibinfo{title}{{Partitioned strong coupling algorithms for
  fluid-structure interaction}},
\newblock \bibinfo{journal}{Computers and Structures} \bibinfo{volume}{81}
  (\bibinfo{year}{2003}) \bibinfo{pages}{805--812}.
\bibitem[{Habchi et~al.(2013)Habchi, Russeil, Bougeard, Harion, Lemenand,
  Ghanem, Valle, and Peerhossaini}]{Habchi2013}
\bibinfo{author}{C.~Habchi}, \bibinfo{author}{S.~Russeil},
  \bibinfo{author}{D.~Bougeard}, \bibinfo{author}{J.~L. Harion},
  \bibinfo{author}{T.~Lemenand}, \bibinfo{author}{A.~Ghanem},
  \bibinfo{author}{D.~D. Valle}, \bibinfo{author}{H.~Peerhossaini},
\newblock \bibinfo{title}{{Partitioned solver for strongly coupled
  fluid-structure interaction}},
\newblock \bibinfo{journal}{Computers and Fluids} \bibinfo{volume}{71}
  (\bibinfo{year}{2013}) \bibinfo{pages}{306--319}.
\bibitem[{Radtke et~al.(2016)Radtke, Larena-Avellaneda, Debus, and
  D{\"{u}}ster}]{Radtke2016}
\bibinfo{author}{L.~Radtke}, \bibinfo{author}{A.~Larena-Avellaneda},
  \bibinfo{author}{E.~S. Debus}, \bibinfo{author}{A.~D{\"{u}}ster},
\newblock \bibinfo{title}{{Convergence acceleration for partitioned simulations
  of the fluid-structure interaction in arteries}},
\newblock \bibinfo{journal}{Computational Mechanics} \bibinfo{volume}{57}
  (\bibinfo{year}{2016}) \bibinfo{pages}{901--920}.
\bibitem[{Verdugo and Wall(2016)}]{Verdugo2015a}
\bibinfo{author}{F.~Verdugo}, \bibinfo{author}{W.~A. Wall},
\newblock \bibinfo{title}{{Unified computational framework for the efficient
  solution of n-field coupled problems with monolithic schemes}},
\newblock \bibinfo{journal}{Computer Methods in Applied Mechanics and
  Engineering} \bibinfo{volume}{310} (\bibinfo{year}{2016})
  \bibinfo{pages}{335--366}.
\bibitem[{Badia et~al.(2008)Badia, Quaini, and Quarteroni}]{Badia2008a}
\bibinfo{author}{S.~Badia}, \bibinfo{author}{A.~Quaini},
  \bibinfo{author}{A.~Quarteroni},
\newblock \bibinfo{title}{{Modular vs. non-modular preconditioners for
  fluid-structure systems with large added-mass effect}},
\newblock \bibinfo{journal}{Computer Methods in Applied Mechanics and
  Engineering} \bibinfo{volume}{197} (\bibinfo{year}{2008})
  \bibinfo{pages}{4216--4232}.
\bibitem[{Houzeaux et~al.(2009)Houzeaux, V{\'{a}}zquez, Aubry, and
  Cela}]{Houzeaux2009}
\bibinfo{author}{G.~Houzeaux}, \bibinfo{author}{M.~V{\'{a}}zquez},
  \bibinfo{author}{R.~Aubry}, \bibinfo{author}{J.~M. Cela},
\newblock \bibinfo{title}{{A massively parallel fractional step solver for
  incompressible flows}},
\newblock \bibinfo{journal}{Journal of Computational Physics}
  \bibinfo{volume}{228} (\bibinfo{year}{2009}) \bibinfo{pages}{6316--6332}.
\bibitem[{Houzeaux et~al.(2011)Houzeaux, Aubry, and
  V{\'{a}}zquez}]{Houzeaux2011}
\bibinfo{author}{G.~Houzeaux}, \bibinfo{author}{R.~Aubry},
  \bibinfo{author}{M.~V{\'{a}}zquez},
\newblock \bibinfo{title}{{Extension of fractional step techniques for
  incompressible flows: The preconditioned Orthomin(1) for the pressure Schur
  complement}},
\newblock \bibinfo{journal}{Computers and Fluids} \bibinfo{volume}{44}
  (\bibinfo{year}{2011}) \bibinfo{pages}{297--313}.
\bibitem[{Casoni et~al.(2015)Casoni, J{\'{e}}rusalem, Samaniego, Eguzkitza,
  Lafortune, Tjahjanto, S{\'{a}}ez, Houzeaux, and V{\'{a}}zquez}]{Casoni2014}
\bibinfo{author}{E.~Casoni}, \bibinfo{author}{A.~J{\'{e}}rusalem},
  \bibinfo{author}{C.~Samaniego}, \bibinfo{author}{B.~Eguzkitza},
  \bibinfo{author}{P.~Lafortune}, \bibinfo{author}{D.~D. Tjahjanto},
  \bibinfo{author}{X.~S{\'{a}}ez}, \bibinfo{author}{G.~Houzeaux},
  \bibinfo{author}{M.~V{\'{a}}zquez},
\newblock \bibinfo{title}{{Alya: Computational Solid Mechanics for
  Supercomputers}},
\newblock \bibinfo{journal}{Archives of Computational Methods in Engineering}
  \bibinfo{volume}{22} (\bibinfo{year}{2015}) \bibinfo{pages}{557--576}.
\bibitem[{Vazquez et~al.(2014)Vazquez, Houzeaux, Koric, Artigues,
  Aguado-Sierra, Aris, Mira, Calmet, Cucchietti, Owen, Taha, and
  Cela}]{Vazquez2014}
\bibinfo{author}{M.~Vazquez}, \bibinfo{author}{G.~Houzeaux},
  \bibinfo{author}{S.~Koric}, \bibinfo{author}{A.~Artigues},
  \bibinfo{author}{J.~Aguado-Sierra}, \bibinfo{author}{R.~Aris},
  \bibinfo{author}{D.~Mira}, \bibinfo{author}{H.~Calmet},
  \bibinfo{author}{F.~Cucchietti}, \bibinfo{author}{H.~Owen},
  \bibinfo{author}{A.~Taha}, \bibinfo{author}{J.~M. Cela},
\newblock \bibinfo{title}{{Alya: Towards Exascale for Engineering Simulation
  Codes}}  (\bibinfo{year}{2014}) \bibinfo{pages}{1--20}.
\bibitem[{F{\"{o}}rster et~al.(2007)F{\"{o}}rster, Wall, and
  Ramm}]{Forster2007}
\bibinfo{author}{C.~F{\"{o}}rster}, \bibinfo{author}{W.~A. Wall},
  \bibinfo{author}{E.~Ramm},
\newblock \bibinfo{title}{{Artificial added mass instabilities in sequential
  staggered coupling of nonlinear structures and incompressible viscous
  flows}},
\newblock \bibinfo{journal}{Computer Methods in Applied Mechanics and
  Engineering} \bibinfo{volume}{196} (\bibinfo{year}{2007})
  \bibinfo{pages}{1278--1293}.
\bibitem[{Causin et~al.(2005)Causin, Gerbeau, and Nobile}]{Causin2005}
\bibinfo{author}{P.~Causin}, \bibinfo{author}{J.~F. Gerbeau},
  \bibinfo{author}{F.~Nobile},
\newblock \bibinfo{title}{{Added-mass effect in the design of partitioned
  algorithms for fluid-structure problems}},
\newblock \bibinfo{journal}{Computer Methods in Applied Mechanics and
  Engineering} \bibinfo{volume}{194} (\bibinfo{year}{2005})
  \bibinfo{pages}{4506--4527}.
\bibitem[{Bungartz et~al.(2015)Bungartz, Lindner, Mehl, and
  Uekermann}]{Bungartz2015}
\bibinfo{author}{H.~J. Bungartz}, \bibinfo{author}{F.~Lindner},
  \bibinfo{author}{M.~Mehl}, \bibinfo{author}{B.~Uekermann},
\newblock \bibinfo{title}{{A plug-and-play coupling approach for parallel
  multi-field simulations}},
\newblock \bibinfo{journal}{Computational Mechanics} \bibinfo{volume}{55}
  (\bibinfo{year}{2015}) \bibinfo{pages}{1119--1129}.
\bibitem[{Uekermann(2016)}]{uekermann2016partitioned}
\bibinfo{author}{B.~W. Uekermann}, \bibinfo{title}{Partitioned fluid-structure
  interaction on massively parallel systems}, Ph.D. thesis, Technische
  Universit{\"a}t M{\"u}nchen, \bibinfo{year}{2016}.
\bibitem[{Haelterman et~al.(2016)Haelterman, Bogaers, Scheufele, Uekermann, and
  Mehl}]{Haelterman2016}
\bibinfo{author}{R.~Haelterman}, \bibinfo{author}{A.~E. Bogaers},
  \bibinfo{author}{K.~Scheufele}, \bibinfo{author}{B.~Uekermann},
  \bibinfo{author}{M.~Mehl},
\newblock \bibinfo{title}{{Improving the performance of the partitioned QN-ILS
  procedure for fluid-structure interaction problems: Filtering}},
\newblock \bibinfo{journal}{Computers and Structures} \bibinfo{volume}{171}
  (\bibinfo{year}{2016}) \bibinfo{pages}{9--17}.
\bibitem[{Mehl et~al.(2016)Mehl, Uekermann, Bijl, Blom, Gatzhammer, and {Van
  Zuijlen}}]{Mehl2016}
\bibinfo{author}{M.~Mehl}, \bibinfo{author}{B.~Uekermann},
  \bibinfo{author}{H.~Bijl}, \bibinfo{author}{D.~Blom},
  \bibinfo{author}{B.~Gatzhammer}, \bibinfo{author}{A.~{Van Zuijlen}},
\newblock \bibinfo{title}{{Parallel coupling numerics for partitioned
  fluid-structure interaction simulations}},
\newblock \bibinfo{journal}{Computers and Mathematics with Applications}
  \bibinfo{volume}{71} (\bibinfo{year}{2016}) \bibinfo{pages}{869--891}.
\bibitem[{Degroote(2013)}]{Degroote2013}
\bibinfo{author}{J.~Degroote},
\newblock \bibinfo{title}{{Partitioned Simulation of Fluid-Structure
  Interaction: Coupling Black-Box Solvers with Quasi-Newton Techniques}}
  \bibinfo{volume}{20} (\bibinfo{year}{2013}) \bibinfo{pages}{185--238}.
\bibitem[{Scheufele(2015)}]{Scheufele2015}
\bibinfo{author}{K.~Scheufele}, \bibinfo{title}{{Robust Quasi-Newton Methods
  for Partitioned Fluid-Structure Simulations}}, Ph.D. thesis,
  \bibinfo{year}{2015}.
\bibitem[{Vierendeels et~al.(2007)Vierendeels, Lanoye, Degroote, and
  Verdonck}]{Vierendeels2007}
\bibinfo{author}{J.~Vierendeels}, \bibinfo{author}{L.~Lanoye},
  \bibinfo{author}{J.~Degroote}, \bibinfo{author}{P.~Verdonck},
\newblock \bibinfo{title}{{Implicit coupling of partitioned fluid-structure
  interaction problems with reduced order models}},
\newblock \bibinfo{journal}{Computers and Structures} \bibinfo{volume}{85}
  (\bibinfo{year}{2007}) \bibinfo{pages}{970--976}.
\bibitem[{Bogaers et~al.(2014)Bogaers, Kok, Reddy, and Franz}]{Bogaers2014}
\bibinfo{author}{A.~E. Bogaers}, \bibinfo{author}{S.~Kok},
  \bibinfo{author}{B.~D. Reddy}, \bibinfo{author}{T.~Franz},
\newblock \bibinfo{title}{{Quasi-Newton methods for implicit black-box FSI
  coupling}},
\newblock \bibinfo{journal}{Computer Methods in Applied Mechanics and
  Engineering} \bibinfo{volume}{279} (\bibinfo{year}{2014})
  \bibinfo{pages}{113--132}.
\bibitem[{Houzeaux et~al.(2018)Houzeaux, Borrell, Cajas, and
  Vázquez}]{HOUZEAUX2018216}
\bibinfo{author}{G.~Houzeaux}, \bibinfo{author}{R.~Borrell},
  \bibinfo{author}{J.~Cajas}, \bibinfo{author}{M.~Vázquez},
\newblock \bibinfo{title}{Extension of the parallel sparse matrix vector
  product (spmv) for the implicit coupling of pdes on non-matching meshes},
\newblock \bibinfo{journal}{Computers and Fluids} \bibinfo{volume}{173}
  (\bibinfo{year}{2018}) \bibinfo{pages}{216 -- 225}.
\bibitem[{Borrell et~al.(2018)Borrell, Cajas, Mira, Taha, Koric, Vázquez, and
  Houzeaux}]{BORRELL2018264}
\bibinfo{author}{R.~Borrell}, \bibinfo{author}{J.~Cajas},
  \bibinfo{author}{D.~Mira}, \bibinfo{author}{A.~Taha},
  \bibinfo{author}{S.~Koric}, \bibinfo{author}{M.~Vázquez},
  \bibinfo{author}{G.~Houzeaux},
\newblock \bibinfo{title}{Parallel mesh partitioning based on space filling
  curves},
\newblock \bibinfo{journal}{Computers and Fluids} \bibinfo{volume}{173}
  (\bibinfo{year}{2018}) \bibinfo{pages}{264 -- 272}.
\bibitem[{Houzeaux and Principe(2008)}]{Houzeaux2008}
\bibinfo{author}{G.~Houzeaux}, \bibinfo{author}{J.~Principe},
\newblock \bibinfo{title}{{A variational subgrid scale model for transient
  incompressible flows}},
\newblock \bibinfo{journal}{International Journal of Computational Fluid
  Dynamics} \bibinfo{volume}{22} (\bibinfo{year}{2008})
  \bibinfo{pages}{135--152}.
\bibitem[{Calderer and Masud(2010)}]{Calderer2010}
\bibinfo{author}{R.~Calderer}, \bibinfo{author}{A.~Masud},
\newblock \bibinfo{title}{{A multiscale stabilized ALE formulation for
  incompressible flows with moving boundaries}},
\newblock \bibinfo{journal}{Computational Mechanics} \bibinfo{volume}{46}
  (\bibinfo{year}{2010}) \bibinfo{pages}{185--197}.
\bibitem[{Turek and Hron(2006)}]{Turek}
\bibinfo{author}{S.~Turek}, \bibinfo{author}{J.~Hron},
\newblock \bibinfo{title}{{Proposal for Numerical Benchmarking of
  Fluid-Structure Interaction between an Elastic Object and Laminar
  Incompressible Flow}},
\newblock \bibinfo{journal}{Fluid-Structure Interaction}
  (\bibinfo{year}{2006}) \bibinfo{pages}{371--385}.
\bibitem[{Cajas et~al.(2015)Cajas, Zavala, Houzeaux, Casoni, V{\'{a}}zquez,
  Moulinec, and Fournier}]{Cajas2015}
\bibinfo{author}{J.~C. Cajas}, \bibinfo{author}{M.~Zavala},
  \bibinfo{author}{G.~Houzeaux}, \bibinfo{author}{E.~Casoni},
  \bibinfo{author}{M.~V{\'{a}}zquez}, \bibinfo{author}{C.~Moulinec},
  \bibinfo{author}{Y.~Fournier},
\newblock \bibinfo{title}{{Fluid structure interaction in HPC multi-code
  coupling}},
\newblock in: \bibinfo{booktitle}{Civil-Comp Proceedings}, volume
  \bibinfo{volume}{107}, \bibinfo{year}{2015}, pp. \bibinfo{pages}{1--26}.
\bibitem[{Degroote et~al.(2008)Degroote, Bruggeman, Haelterman, and
  Vierendeels}]{Degroote2008}
\bibinfo{author}{J.~Degroote}, \bibinfo{author}{P.~Bruggeman},
  \bibinfo{author}{R.~Haelterman}, \bibinfo{author}{J.~Vierendeels},
\newblock \bibinfo{title}{{Stability of a coupling technique for partitioned
  solvers in FSI applications}},
\newblock \bibinfo{journal}{Computers and Structures} \bibinfo{volume}{86}
  (\bibinfo{year}{2008}) \bibinfo{pages}{2224--2234}.
\bibitem[{Santiago et~al.(2018)Santiago, Aguado-Sierra, Zavala-Ak{\'{e}},
  Doste-Beltran, G{\'{o}}mez, Ar{\'{i}}s, Cajas, Casoni, and
  V{\'{a}}zquez}]{Santiago2018}
\bibinfo{author}{A.~Santiago}, \bibinfo{author}{J.~Aguado-Sierra},
  \bibinfo{author}{M.~Zavala-Ak{\'{e}}}, \bibinfo{author}{R.~Doste-Beltran},
  \bibinfo{author}{S.~G{\'{o}}mez}, \bibinfo{author}{R.~Ar{\'{i}}s},
  \bibinfo{author}{J.~C. Cajas}, \bibinfo{author}{E.~Casoni},
  \bibinfo{author}{M.~V{\'{a}}zquez},
\newblock \bibinfo{title}{{Fully coupled fluid-electro-mechanical model of the
  human heart for supercomputers}},
\newblock \bibinfo{journal}{International Journal for Numerical Methods in
  Biomedical Engineering} \bibinfo{volume}{34} (\bibinfo{year}{2018}).

\end{thebibliography}

\end{document}